%% file: main.tex
\documentclass[sigconf,authorversion,nonacm]{acmart}

\usepackage{graphicx}
\usepackage{textcomp}
\usepackage{xcolor}

\usepackage{times}
\usepackage[english]{babel}
\usepackage{blindtext}
\usepackage{graphics} 
\usepackage{amsmath}
\usepackage{amsfonts}
\usepackage{enumitem}

\usepackage[noend]{algpseudocode}
\usepackage[font=small,skip=4pt]{caption}
\usepackage{comment}
\usepackage[font=footnotesize]{subcaption}
\usepackage[linesnumbered,algoruled,boxed,lined]{algorithm2e}
\usepackage{xspace}

\newcommand{\name}{GreenNFV\xspace}

\newcommand{\cut}[1]{}

\AtBeginDocument{%
  }

\begin{document}

\title{\name: Energy-Efficient Network Function Virtualization with Service Level Agreement Constraints}

\author{MD S Q Zulkar Nine}
\orcid{0000-0001-6218-0199}
\affiliation{%
  \institution{University at Buffalo}
  \city{Buffalo}
  \state{New York}
  \country{USA}}
\email{mdsqzulk@buffalo.edu}

\author{Tevfik Kosar}
\affiliation{%
  \institution{University at Buffalo}
  \city{Buffalo}
  \state{New York}
  \country{USA}}
\email{tkosar@buffalo.edu}

\author{Fatih Bulut}
\affiliation{%
  \institution{IBM Research}
  \city{Yorktwon Heights}
  \state{New York}
  \country{USA}}
\email{mfbulut@us.ibm.com}

\author{Jinho Hwang}
\affiliation{%
  \institution{AI Infra R\&D, Meta}
  \city{New York City}
  \state{New York}
  \country{USA}}
\email{jinhohwang@meta.com}

\renewcommand{\shortauthors}{Zulkar et al.}

\begin{abstract}
Network Function Virtualization (NFV) platforms consume significant energy, introducing high operational costs in edge and data centers. This paper presents a novel framework called GreenNFV that optimizes resource usage for network function chains using deep reinforcement learning. GreenNFV optimizes resource parameters such as CPU sharing ratio, CPU frequency scaling, last-level cache (LLC) allocation, DMA buffer size, and packet batch size. GreenNFV learns the resource scheduling model from the benchmark experiments and takes Service Level Agreements (SLAs) into account to optimize resource usage models based on the different throughput and energy consumption requirements. Our evaluation shows that GreenNFV models achieve high transfer throughput and low energy consumption while satisfying various SLA constraints. Specifically, GreenNFV with {\em Throughput SLA} can achieve $4.4\times$ higher throughput and $1.5\times$ better energy efficiency over the baseline settings, whereas GreenNFV with {\em Energy SLA} can achieve $3\times$ higher throughput while reducing energy consumption by 50\%.
\end{abstract}




\maketitle

\input{section-introduction.tex}

\input{section-relatedwork.tex}

\input{section-microbechmarking.tex}
\input{section-modeldescription.tex}

\input{section-Implementation.tex}
\input{section-evaluation.tex}

\input{section-conclusion.tex}

\section*{Acknowledgments}
This project is in part sponsored by the National Science Foundation (NSF) under award numbers CCF-2007829 and OAC-1842054.

\bibliographystyle{ACM-Reference-Format}
\bibliography{main}

\end{document}

%% file: section-introduction.tex
\section{Introduction}
\label{sec:introduction}

Network Function Virtualization (NFV) \emph{virtually} decouples network functions (e.g., firewalls, routers, tunneling gateways, CDNs) from the physical devices, enabling the deployment of new network services with increased agility and faster time-to-value. NFVs bring the benefits of cloud computing and virtualization to the networking domain, facilitating virtualized network functions (VNFs) to be implemented in software and run on any physical server~\cite{kar2017energy}. 
Telecommunication service providers (TSPs) prefer NFV over purchasing, storing, and operating physical equipment since it not only reduces their capital and operating expenditures, but also provides them with more flexibility to open up their network capabilities and services to users, and the ability to deploy and support new network services faster and cheaper~\cite{mijumbi2015network}. Since energy bills represent more than 10\% of TSPs’ operating expenses~\cite{GWATT}, reducing energy consumption while sustaining throughput is one of the strong selling points of NFV as well as an open area of research for improvement. 

The NFV platforms mainly focus on high throughput and low latency, as the virtualization overhead of VNFs has been the primary deterrent to its adoption in lieu of hardware-based middleboxes. VNFs using Intel Data Plane Development Kit (DPDK) can achieve more than 100 Gbps throughput at the cost of high CPU power~\cite{linhao:dpdk}. The poll mode driver in DPDK uses complete cycles of dedicated cores to maximize the packet processing throughput. Despite the line-rate throughput in DPDK, it significantly adds up high energy consumption to the data center network. There is no systematic control to reduce the energy consumption in these packages. Moreover, existing work in this area fails to achieve high data transfer throughput and high energy efficiency simultaneously.

In this paper, we present a novel resource scheduling framework, \name, to achieve energy-efficient network function virtualization under various Service Level Agreement (SLA) constraints. \name uses Deep Reinforcement Learning and allows a customizable energy efficiency rate for the target data transfer throughput. \name makes it easy to dynamically control energy consumption rate given the target transfer throughput through resource controls such as CPU allocation, CPU frequency scaling, last-level cache (LLC) allocation, DMA buffer size, and packet batch size. 

\name harmonically controls hardware (HW) components. Statistical analysis of the network flows enables \name to identify packet arrival rates and traffic patterns. The packet arrival rate decides the polling frequency to match enough resources to achieve the target performance. 

Compute resources are scheduled to those network functions (NFs) according to network load using the HW control knobs. 

The main contributions of this paper include:

\begin{enumerate}[label=(\arabic*)]

\item Introduction of \name, a highly effective resource scheduling framework for energy-efficient network function virtualization under different SLA constraints;

\item Translation of resource scheduling problem into deep deterministic policy gradient (DDPG) algorithm, a value-based actor-critic reinforcement learning algorithm, which is very effective for continuous (real-valued) and high-dimensional action space;

\item Presentation of three novel resource optimization models based on different energy-aware service level agreements (SLAs), which enable the TSPs to minimize energy consumption without compromising the performance guarantees given to the customers; and

\item Development of a distributed learning model by extending the concept of prioritized experience replay to learn from multiple workers simultaneously.

\end{enumerate}

We have implemented \name on top of the NetVM platform~\cite{179739} and devised a mix of callback and polling to better control the network scheduling. The learned models are validated with various SLA constraints (such as Maximum Throughput, Minimum Energy, and Energy Efficiency). Our evaluation shows that \name can achieve both high throughput and low energy at the same time while meeting the SLA constraints. \name 
with {\em Maximum Throughput SLA} can achieve 4.4$\times$ throughput improvement over the baseline (without any optimization) while consuming 33\% less energy. \name with {\em Minimum Energy SLA} can achieve $3\times$ throughput improvement of the baseline while reducing energy consumption by half.

The rest of the paper is organized as follows: $\S$\ref{sec:background} provides background information and discusses the related work in this area; $\S$\ref{sec:benchmark} presents our benchmark study on the individual impact of each resource on system performance and energy consumption; $\S$\ref{sec:modeldesign} presents \name model design and implementation considerations; 
$\S$\ref{sec:evaluation} elaborates the evaluation of \name; and $\S$\ref{sec:conclusion} concludes the paper with a discussion on future work.

%% file: section-relatedwork.tex
\section{Background and Related Work}
\label{sec:background}

VNF placement for improved resource utilization is a well-studied problem in the literature. 
The goal of efficient VNF placement is to minimize the inter-core or processor communication, thus reducing the cache eviction and memory accesses. In particular, as service chains process the same packets, the placement can efficiently group these chains in the same core and processor to achieve higher performance and lower energy consumption. The packets can also be processed in parallel and share the common functionalities in the protocol level~\cite{Liu:2018:MHP:3230543.3230563}. 
A substantial body of literature delves into the optimization of end-to-end data transfer throughput, encompassing the tuning of various application-layer and kernel-layer parameters~\cite{alan2014energy, kosar2005data, balman2009dynamic, kim2015highly, deelman2006makes, balman2007data}.
However, this body of work has not addressed this optimization within the context of VNFs.

Kulkarni et al.~\cite{Kulkarni:2017} address the placement of middleboxes and VNFs for a performance target or efficient resource usage.
Bari et al.~\cite{bari2015orchestrating} present an Integer Linear Programming (ILP) model for VNF orchestration.
Marotta et al.~\cite{marotta2017fast} propose mixed-integer optimization with online heuristics for the VNF placement problem under resource demand uncertainty.
Kaur et al.~\cite{kaur2019energy} leverage evolutionary optimization algorithms to solve the VNF deployment problem in a multi-domain software-defined networking (SDN) setup. 
Wang et al.~\cite{wang2016joint} propose a model to jointly optimize NFV resource allocation in three phases: virtual network functions (VNFs) chain composition, VNFs forwarding graph embedding, and VNFs scheduling. They apply a general cost model to consider network costs and service performance. 
Qu et al.~\cite{qu2016delay} consider VNF transmission and processing delays and formulate the joint problem of VNF scheduling and traffic steering as a mixed-integer linear program (MILP) to minimize the makespan/latency of the overall VNFs’ schedule.
These works do not consider the energy consumption aspect of the VNF placement problem.
Khoury et al.~\cite{el2016energy} formulate an ILP problem to schedule network flows into virtual network functions so that the model can reduce power consumption. 
Al-Quzweeni et al.~\cite{al2019optimized} employ the MILP optimization model to minimize total power consumption in the context of 5G networks by optimizing the VM location and VM server utilization. However, these models only work with the offline version of the problem and fail to adapt to the dynamically changing network conditions.

\begin{figure*}[t]
\begin{centering}
\begin{subfigure}[t]{0.32\textwidth}
        \centering
        \includegraphics[keepaspectratio=true,width=54mm]{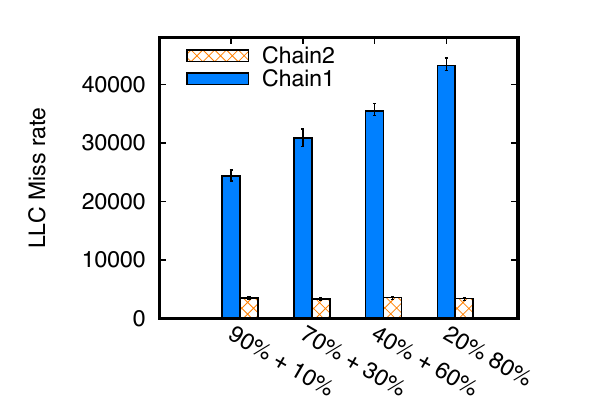}
        \caption{LLC miss rate}
\end{subfigure}
\begin{subfigure}[t]{0.32\textwidth}
    \centering
        \includegraphics[keepaspectratio=true,width=54mm]{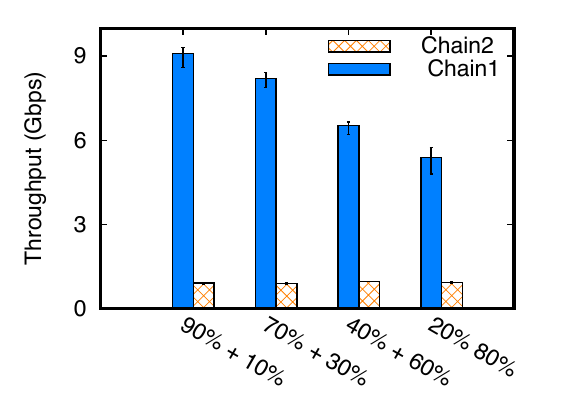}
        \caption{NF achieved throughput}
\end{subfigure}
\begin{subfigure}[t]{0.32\textwidth}
    \centering
        \includegraphics[keepaspectratio=true,width=54mm]{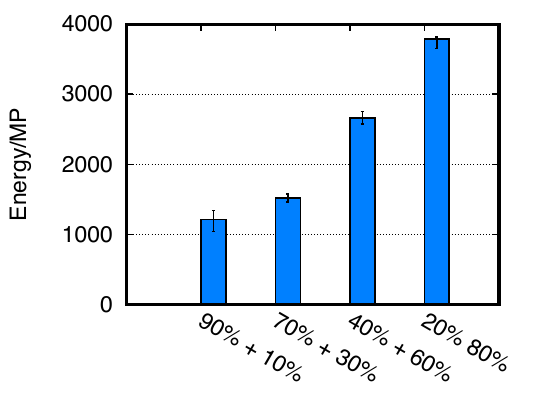}
        \caption{Energy consumption}
\end{subfigure}
\caption{Micro-benchmarking of LLC size: effect of LLC on NF throughput and energy consumption.}
\label{fig:benchmark_llc}
\end{centering}
\end{figure*}

Numerous works have been done to address the NFV resource allocation problem.
Chen~\cite{chen2020energy} studies NFV resource allocation in edge computing environments using a genetic algorithm-based approach. 
However, it comes with a assumption that the energy consumption can be measured before placement. In many cases, it is not realistic. 
Sun et al.~\cite{sun2020energy} propose an energy-aware routing algorithm that considers network function chains' traffic requirements and bandwidth consumption to minimize server energy consumption.
Marotta and Kassler~\cite{marotta2016power} present a joint resource and flow routing assignment problem for VNF placement to minimize the power consumption of the servers and switches needed to deploy the overall virtualized infrastructure and the routing graph. 
Kar et al.~\cite{kar2017energy} design a dynamic energy-saving model with NFV technology using an M/M/c queuing network with the minimum capacity policy where a certain amount of load is required to start the machine, which increases the utilization and avoids frequent changes of the device states. They formulate an energy-cost optimization problem with capacity and delay as constraints and propose a heuristic solution for the dynamic placement of VNF chains to solve this NP-hard problem. 
Zhang et al.~\cite{zhang2019near} propose a polynomial near-optimal algorithm based on the Markov approximation technique to effectively place VNFs at different locations and steer service function chain requests while minimizing energy consumption. 
Iqbal et al.~\cite{faisal12} propose a scheme that uses P and C-state of the processor to reduce both active and idle power consumption. They used ANN for traffic prediction. 
Sun et al.~\cite{sun2020energy} studies the online deployment of the service function chain and proposed energy-aware routing and adaptive delay shutdown mechanism. They used a simulated environment to verify the approaches. 
Pei et al.~\cite{Pei2020} proposed a VNF placement algorithm based on Double Deep Q Network and threshold-based policy. The solution considers the transfer throughput, end-to-end delay, running time, and load balancing. 

\name considers a range of system-level control knobs (e.g., number of CPU cores, CPU frequency scaling, last-level-cache (LLC) allocation, DMA buffer size, and packet batch size). It also consolidates the VNFs based on the flow path and minimizes the cache eviction, reducing memory access and increasing CPU utilization.
The resource allocation for VNFs decides performance and energy efficiency. CPU allocation is controlled by frequency scaling and turning off idle CPU cores.
DMA buffer size also plays a vital role in achieving high performance in the NFV environment, as it makes the best use of Intel Data Direct I/O to avoid cache eviction~\cite{amin2018:resQ}. 
%
Cache Allocation Technology~\cite{IntelCAT} provides an OS-level control of partitioning the LLC, where LLC is shared among multiple cores and renders fewer memory accesses, higher performance, and lower energy consumption. \name essentially learns how to schedule these hardware resources and applies new configurations in run-time for service chains. Such ability enables the system to adapt when network conditions change and ensures high packet processing throughput and energy efficiency.

%% file: section-microbechmarking.tex
\section{Resource Impact Analysis}
\label{sec:benchmark}

NFs run on a virtualized machine, sharing hardware resources such as CPU cores, cache, memory, and network queues. The achievable throughput of NFs could differ depending on how the resources are configured for NFs. 

\name leverages a mix of controllable resources to optimize resource usage towards throughput and energy efficiency.

\noindent {\bf Last-level cache (LLC)} is an essential element for high-speed network performance~\cite{amin2018:resQ,179739}. NFs in a service chain access the same packet, and caching the packet suppresses memory access. Intel CAT can provide finer-grained control over the shared LLC. 
As an advanced cache system, a modern Intel processor has Data Direct I/O (DDIO) technology~\cite{IntelDDIO}, 10\% of the LLC is allocated for DDIO and used for packet processing only. In legacy systems, NIC directly writes packets into the main memory, and then the CPU reads packets from the main memory to LLC. Instead of such direct implementation, NIC can write the packet directly to the LLC. 
We perform micro-benchmarking on two network function chains named - C1 and C2. The input flows in these function chains are 13 Mpps and 1 Mpps, respectively. We schedule different portions of LLC to these chains and record their impact. Assume the tuple, $(x,y)_{llc}$, where, $x$ and $y$ are LLC allocation for C1 and C2 respectively, we consider 4 allocations [(90\%,10\%), (70\%,30\%), (40\%,60\%), (20\%,80\%)] and record each throughput and energy consumption. 
Figure~\ref{fig:benchmark_llc}(a) shows the cache miss rates, and Figure~\ref{fig:benchmark_llc}(b) shows the achieved throughput of the function chains. The (90\%,10\%) allocation is reasonable since it allocates LLC proportional to the input flows. The performance of C1 starts dropping as we reduce the LLC allocation for C1 and increase it for C2. Figure~\ref{fig:benchmark_llc}(c) shows the energy consumption of the same cases. We see that decreasing the LLC for C1 can increase energy consumption significantly. This is due to the frequent cache misses in small LLC sizes. Therefore, optimal LLC allocation to many different network function chains can reduce packet processing energy consumption. 

\begin{figure}[t]
\centering
\includegraphics[keepaspectratio=true,angle=0,width=80mm]{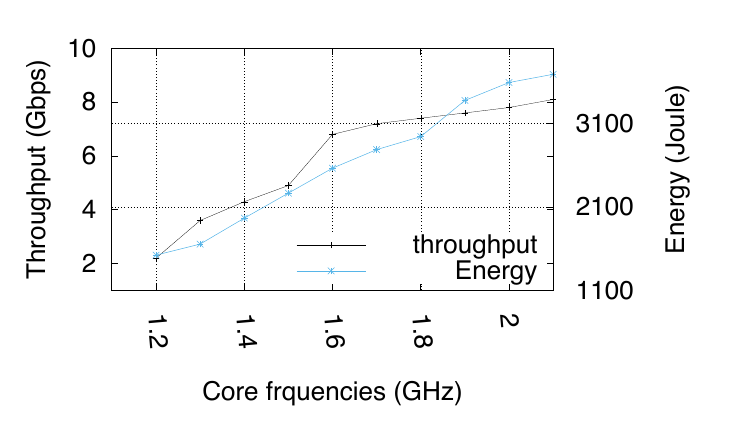}
\caption{Micro-benchmarking of CPU frequencies: effect of CPU frequencies on NF throughput and energy efficiency.} 
\label{fig:benchmark_CPU_frequencies}
\end{figure}

\begin{figure*}[t]
\begin{centering}
\hspace{6mm}
\begin{subfigure}[t]{0.45\textwidth}
    \centering
        \includegraphics[keepaspectratio=true,width=63mm]{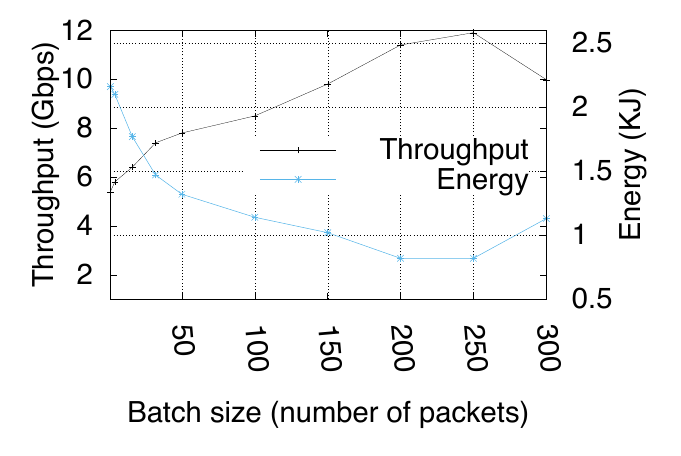}
        \caption{NF achievable throughput}
\end{subfigure}
\begin{subfigure}[t]{0.45\textwidth}
        \hspace{10mm}
        \includegraphics[keepaspectratio=true,width=55mm]{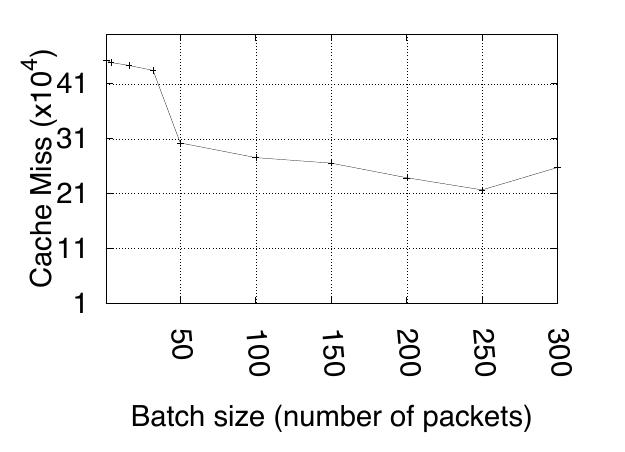}
        \caption{LLC cache miss rate}
\end{subfigure}
 \caption{Micro-benchmarking of batching size: effect of batch size on NF throughput and energy efficiency.}
 \label{fig:benchmark_batch_size}
 \end{centering}
 \end{figure*}

\begin{figure*}[t]
\begin{centering}
\begin{subfigure}[t]{0.45\textwidth}
    \centering
        \includegraphics[keepaspectratio=true,width=55mm]{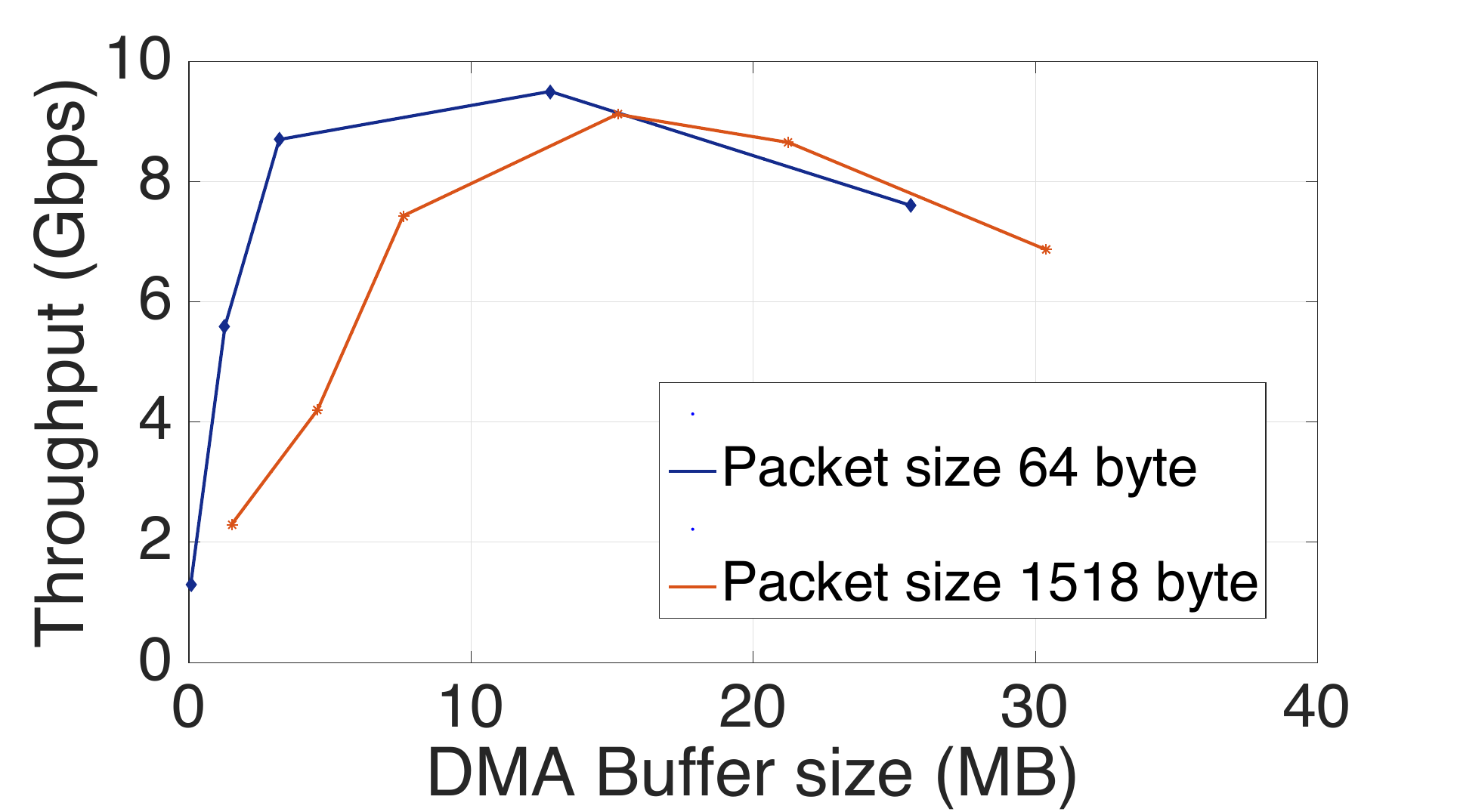}
        \caption{NF achieved throughput}
\end{subfigure}
\begin{subfigure}[t]{0.45\textwidth}
    \centering
        \includegraphics[keepaspectratio=true,width=55mm]{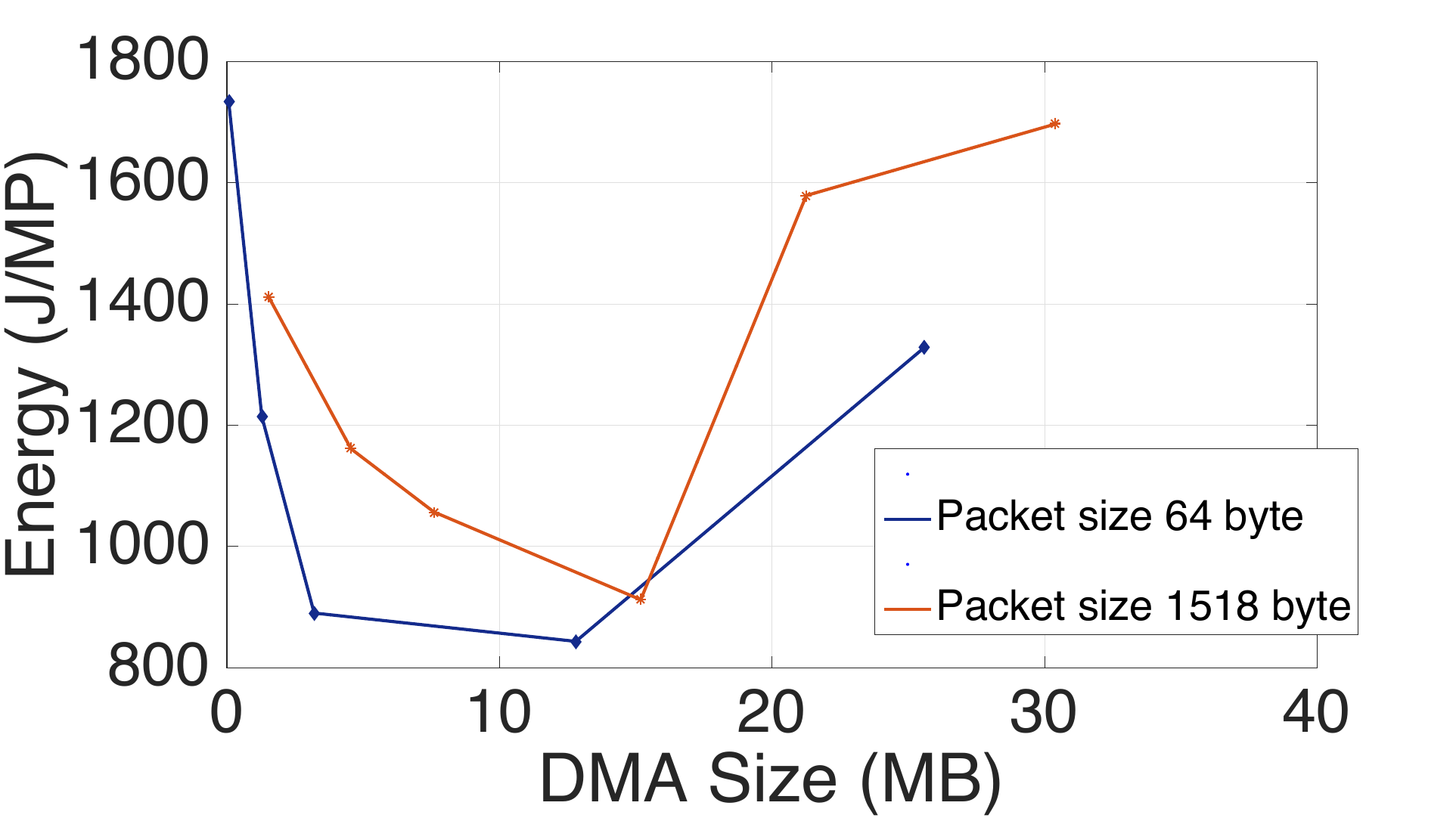}
        \caption{NF energy consumption}
\end{subfigure}
\caption{Micro-benchmarking of DMA buffer size: effect of DMA buffer size on NF throughput and energy efficiency.}
\label{fig:benchmark_dma}
\end{centering}
\end{figure*}

 \noindent {\bf Dynamic frequency scaling of CPUs} is one of the major control knobs in energy efficiency. 
 Low frequency can hamper the throughput, while a high value can consume extra energy. We performed micro-benchmarking on the NFV system with different levels of CPU frequencies. We used an Intel Xeon E5 (V4) Processor with 64 GB main memory and an NF chain of three NFs. The line rate traffic with a large packet size (1518 Bytes) is fed into the function chain. Figure~\ref{fig:benchmark_CPU_frequencies} shows the impact of CPU frequency scaling on the throughput of the network packet processing. The packet processing rate and energy consumption increase when we increase the CPU frequencies. However, the growth is non-linear. The result varies when different network chains with different NFs are used.

\noindent {\bf Batching} enables the packets to be processed in groups, which provides improved cache locality to the packets. NFs are instructions that need to be performed over many subsequent packets. Batching the packets provides faster access to the packets. Fetching the packets from the main memory is expensive. Instead of fetching each packet for processing, a batch of packets can be fetched at the same time as fetching a single packet. However, the size is crucial because excessive batching can overload the LLC and increase the miss rate. 
Figure~\ref{fig:benchmark_batch_size} shows the impact of batch size on throughput, energy consumption, and cache miss rates. The increase in batch size can increase the packet processing speed of the NF chain, reducing the cache miss rate to a certain level. Afterward, the throughput decreases, and the cache miss rate and energy consumption increase. The optimal level of batch size also depends on the network function and the chain.

\noindent {\bf DMA buffer size} needs to align with LLC because the large DMA buffer size can overflow LLC size, so the LLC miss rate can increase due to cache eviction.
Figure~\ref{fig:benchmark_dma} shows a network function's packet processing speed and energy consumption and compares the result of processing two flows with different packet sizes. 
We can see that the DMA buffer size can control the packet processing speed and the energy consumption of the NFs. DMA buffer size can steadily increase the performance up to a certain level. During this time, energy consumption tends to decrease. With increased throughput, the system can process quickly, reducing the idle energy consumption of the system.

%% file: section-modeldescription.tex
\section{\name Design}
\label{sec:modeldesign}

This section discusses the design of the models that \name uses in its architecture. 

\subsection{Design Considerations}
\label{subsec:design_considerations}

Controlling and distributing the resources optimally among NF functions are the main objectives of \name. For example, when CPU cores are shared, \name needs to appropriately distribute the CPU time and control the CPU frequency level. As discussed in the previous section, in addition to CPU, \name also considers LLC allocation, network buffer size, packet batch size, and packet prefetching. \name aims to provide the optimal amount of resources to the NF chains to achieve the most efficient resource utilization according to the energy constraints. Different chains may require different Quality of Service (QoS). From a telecommunications service provider's standpoint, it is essential to support each scenario and establish a Service Layer Agreement (SLA) with the client for each of the chains. \name considers the following SLAs.

\noindent \textbf{Energy SLA:} \name optimizes resources for specific energy constraints. Then it tries to optimize performance under the defined energy constraint. This scenario is ideal for chains that have specific energy budgets while maximizing the throughput, $T_{f_i}$ for each flow, $f_i$. Equation \ref{eq:energy_sla} shows such optimization.

\begin{equation}
\begin{array}{ll}
\underset{ \{resources\} }{\mathrm{argmax}} & \psi_{T} = \displaystyle\sum_{i=1}^{n} T_{f_i} \\
\text{subject to.} & E \leq \mathbb{E}_{SLA} \\ 
\end{array}
\label{eq:energy_sla}
\end{equation}

\noindent \textbf{Throughput SLA:} Some flows may require strict throughput guarantees. Hence, the administrators may want to guarantee a certain throughput level as defined in the SLA while trying to minimize energy consumption. Equation \ref{eq:throughput_sla} shows such optimization. 

\begin{equation}
    \begin{array}{ll}
        \underset{ \{resources\} }{\mathrm{argmin}} & \psi_{E} =  \displaystyle\sum_{i=1}^{n} E_{f_i} \\
                                \text{subject to.} & T \geq \mathbb{T}_{SLA} \\ 
    \end{array}
\label{eq:throughput_sla}
\end{equation}
Here, $E_{fi}$ is the energy consumption of each NF in the NF chain. 

\vspace{1mm}
\noindent \textbf{Energy Efficiency SLA:} This SLA strictly aims to optimize the energy efficiency of the system. We define energy efficiency as the number of bytes transferred in unit time using a unit amount of energy. Therefore, when a system processes $B$ bytes of packets in $t_p$ time and spends $E$ amount of energy, then the goal of the SLA is to increase the energy efficiency, $\lambda$ of the system that is explained as,

\begin{equation}
    \begin{array}{ll}
        \underset{ \{resources\} }{\mathrm{argmax}}  \lambda = \dfrac{B}{E\times t_p} = \dfrac{T}{E},  \\
    \end{array}
\label{eq:energy_efficiency_sla}
\end{equation}

\noindent where $T$ is the throughput of the system. 

In this work, we used a nonlinear power model~\cite{Fal2007power} to estimate the power consumption of the CPU as,
\begin{equation}
    P_u = (P_{max} - P_{idle}) (2u-u^{h}) + P_{idle}
\end{equation}

Where $P_{idle}$ and $P_{max}$ are the average power consumption of the idle server and the average power consumption of the fully utilized server. $u$ is the CPU utilization and $h$ is the calibration parameter. We used the Yokogawa WT210 power meter to measure the actual power to validate the model and compute $h$.

\subsection{Baseline Heuristic Approach}
\label{subsec:heuristic}

Initially, we propose a baseline approach for compute-resource scheduling in the NFV environment. The steps are listed in Algorithm~\ref{algo:heuristics}. The algorithm initially assigns a fixed number of cores to the NFs with a predefined frequency value. Then it computes LLC size, DMA buffer size, and the batch size, as shown in \textit{Line (4-6)} of Algorithm \ref{algo:heuristics}. Then it periodically checks the constraint and dynamically updates the number of cores, core frequencies, and batch size.

\begin{algorithm}[h]
\small
    \SetKwData{CoreFreq}{core\_frequency}
    \SetKwData{Cores}{cores}
    \SetKwData{Batch}{batch\_size}
    \SetKwData{LLCsize}{LLC\_size}
    \SetKwData{DMAbufsize}{DMA\_buffer\_size}
    \SetKwData{packetsize}{packet\_size}
    \SetKwData{energyefficiency}{energy\_efficiency}
    \SetKwData{throughput}{throughput}
    \SetKwData{energyconsumed}{energy\_consumed}
    \SetKwData{batchsize}{batch\_size}
	\BlankLine
	Allocate cores and frequencies evenly to each NF \\
	\Cores $\leftarrow$ 1\\
	\CoreFreq[1:\Cores] $\leftarrow$ median(\CoreFreq)  \\
	\Batch $\leftarrow$ 2 \\
	\LLCsize $\leftarrow$ proportion to flow rate \\
	\DMAbufsize $\leftarrow$ $\dfrac{\LLCsize }{\packetsize \times \batchsize}$\\
	
	Periodically - check the throughput and energy consumption\\
        \tcp{\scriptsize $\lambda$ is energy efficiency} 
        
        $\lambda$ $\leftarrow$ $\dfrac{\throughput }{\energyconsumed}$\\
	\uIf{$\lambda$ < threshold1 }{ 
	    Select nearest smaller  \CoreFreq that is available
	}
	\uElse{
	    Select nearest larger  \CoreFreq that is available
	}
        
	\uIf{$\lambda$ < threshold2 }{ 
	    \Batch $\leftarrow$ \Batch + 1 
	}
	\uElse{
	    \Batch $\leftarrow$ \Batch -- 1 
	}
\caption{Baseline Heuristics Algorithm}
\label{algo:heuristics}
\end{algorithm}

This simplistic algorithm has many issues that can be addressed efficiently. There is a diverse range of network functions - CPU intensive, memory-intensive, lightweight process (e.g., NAT, firewall), and more heavyweight (e.g., Evolved Packet Core). These NFs can have distinct resource requirements. Moreover, network flows can be highly dynamic. Resource allocation is also highly dependent on flow characteristics. All these resources have a non-linear effect on each other. 
Therefore, such a simplistic resource allocation is inefficient in scheduling resources. A learning-based model can achieve significant improvement over the heuristic-based approach.

\subsection{\name Deep Learning Model}
\label{subsec:rl}

The heuristic approach works well for some predefined scenarios; 

However, this approach cannot adapt to the dynamic nature and varying requirements of the different kinds of flows. 
To provide fine-tuned control over performance and energy consumption, we need a model that can fine-tune resources under various known and unknown scenarios. It needs to consider network condition changes over time and should have the ability to adapt new parameter configurations in real time. 
Therefore, the most suitable solution for this problem is a class of algorithms that can train from historical data and perform dynamic resource allocation based on the nature of the flow and available resources. 
In particular, for our problem in hand, \name considers Reinforcement Learning (RL), a sub-field of machine learning. 

RL algorithms learn from the environment and get experienced over time through learning. It can adapt its decisions based on changing environmental conditions. For this problem, an RL-based solution is most suitable to control the system parameters to achieve the performance and energy requirements simultaneously while satisfying the SLA constraints. 
Q Learning~\cite{watkins1992} is a basic RL approach. It uses a (state, action) table of Q values to determine the best action. The algorithm takes discrete values of states and actions. Any continuous value state or action needs to be appropriately discretized. The state space is the set of all the flows along with their throughput, energy consumption, flow completion time, and flow size. The problem arises when we try to discretize the action and state spaces. The number of discrete actions grows exponentially with respect to the discretization level. For example, we have five different actions - number of cores, LLC size, DMA buffer size, batch size, and CPU core frequencies. When we choose $k$ discrete levels for each action, the number of actions becomes $O(k^5)$. For $n$ number of flows, the action space becomes $O(n \times k^5)$. Q-learning requires many entries in the Q-table, and it becomes highly inefficient to maintain such a large table. 

In \name, we model the achievable NF transfer throughput and energy consumption with deep neural networks (DNNs) and use a distributed training framework to ensure efficient computing time for analyzing large historical logs. 
An effective algorithm used to train DNNs in a distributed setting is Distributed Deep Q-Learning~\cite{ong2015distributed}. %
Deep Q-networks (DQNs)~\cite{mnih2015human} are convolutional neural networks trained with a variant of Q-learning~\cite{watkins1992}. DQNs do not require a large Q-table. Instead, they use Deep Learning (DL) to learn the table themselves. DQN has proven very efficient for some application areas, such as Arcade gaming.
DQN introduced several techniques to stabilize learning, such as experience replay, target network, and clipping rewards. However, DQN cannot process a high number of actions in continuous space. Because of the DNN, the output layer can only handle a handful of actions. 

To make the learning process efficient and robust among many machines, we need a distributed learning framework to learn from machines in parallel. It also needs to process ample state and action space efficiently. For this reason, we choose Deep Deterministic Policy Gradient (\emph{DDPG})~\cite{Lillicrap15} that can process the continuous state and action space.  
DDPG is very effective for continuous (real-valued) and high-dimensional action spaces. DDPG uses a stochastic behavior policy for search space exploration but estimates a deterministic target policy that is much easier to learn. This provides us with both highly accurate and fast-converging predictive modeling for throughput optimization. 
DDPG can directly parameterize the policy function and optimize its value. On top of DDPG, to perform learning in distributed settings, \name utilizes the Ape-X framework~\cite{horgan2018distributed}. Below, we explain the use of DDPG and Ape-X in \name in detail.

\subsubsection{ Deep Reinforcement Learning Formulation} 

In a standard reinforcement learning framework, an agent interacts with the environment in discrete time. In each time step $t$, the agent observes the state $x_t \in \mathcal{X}$ and chooses an action $a_t \in \mathcal{A}$. When the state changes to $x_{t+1}$, the agent receives reward $r(x_t, a_t) \in \mathbb{R}$. 
The agent's behavior is controlled by policy $\pi : \mathcal{X} \rightarrow \mathcal{A}$ that translates states to actions. The state-action value function's expected return values can be defined as, $Exp[\sum_{t=0}^{\infty}\gamma^{t}r_{t}(x_t, a_t)]$, where $\gamma_t \in (0,1]$ is the discount factor. 

The agent takes actions based on the policy, $\pi(a|x)$; however, in most cases, it is almost impossible to store all possible (state, action) pairs. A function approximator, $\pi_{\theta}(a|x)$ with parameter vector, $\theta$ can be used to learn the policy function. The deep deterministic policy gradient was derived in \cite{silver14DDPG}. Given a deterministic policy, $\mu_{\theta} : \mathcal{X} \rightarrow \mathcal{A}$ that can approximate optimal action value for a given state and the discounted state distribution (i.e., the probability of visiting the state in the future), $\rho^{\pi}$,  the performance objective can be written as, 

\begin{equation}
    \begin{split}
        J(\rho_{\theta}) & = \int_{\mathcal{S}} \rho^{\mu}(x) \int_{\mathcal{A}} r(x,\mu_{\theta}(x))dadx \\
                         & = Exp_{x\sim\rho^{\mu}}[r(x,\mu_{\theta}(x))]
    \end{split}
\end{equation}

Silver et al.~\cite{silver14DDPG} proves that the gradient of the objective can be written as, 

\begin{equation}
        \nabla_{\theta} J(\rho_{\theta}) = \mathbb{E}_{x\sim\rho^{\mu}}[\nabla_{\theta} \mu_{\theta}(x) \nabla_{a} Q^{\mu} (x,a) |_{a=\mu_{\theta}(x)}]
\end{equation}

DDPG~\cite{silver14DDPG} is an actor-critic model that initializes critic network $Q_{\theta}(s,a)$ and a actor-network, $\mu_{\theta}(x|\theta_\mu)$ with weights, $\theta_{Q}$  and $\theta_{\mu}$. These two networks are trained with a mini-batch of samples. The samples are the transitioning tuples $(x_i,a_i,r_i,x_{i+1})$ when an agent observe $x_i$ state, performs $a_i$ action to receives $r_i$ reward and change the state to $x_{i+1}$. The networks are trained on random samples from the experience replay buffer so that the divergence due to the correlation among the samples is minimized. 
It also initializes the target network, $Q'$, and $\mu'$ to update the actor-critic network smoothly. State update rules are presented in the Algorithm~\ref{algo:DDPG}. More details of the algorithm can be found in~\cite{Lillicrap15}.

\begin{algorithm}[t]
\small
	\BlankLine
    Select action, $a_t = \mu_{\theta^{\mu}}(x) + \mathcal{N}_t$ \\
    	Store transition, $(x_i,a_i,r_i,x_{i+1})$ in Replay buffer, $R$ \\
    	Select random minibatch, $N$ transitions from $R$ \\
    	Set, $y_i = r_i + \gamma Q_{\theta^{Q'}}^{'}(x_{i+1}, \mu_{\theta_{\mu'}}^{'} (x_{i+1}))$ \\
    	Update critic by minimizing the loss: \\
    	$L = 1/N \sum_i (y_i - Q_{\theta^{Q}} (s_i,a_i))^{2}$ \\
    	Update actor using the sampled policy gradient: \\
    	Update the target networks: \\
    	$\theta^{Q'} \leftarrow \tau \theta^{Q} + (1-\tau) \theta^{Q'}$\\
    	$\theta^{\mu'} \leftarrow \tau \theta^{\mu} + (1-\tau) \theta^{\mu'}$ 
\caption{DDPG Algorithm~\cite{Lillicrap15}}
\label{algo:DDPG}
\end{algorithm}

\begin{algorithm}[t]
\small
	\SetKwFunction{Actor}{NF\_CONTROLLER}
	\SetKwFunction{remotecall}{REMOTE\_CALL}
	\SetKwFunction{getStat}{COLLECT\_STATE}
	\SetKwFunction{CentralLearner}{CENTRAL\_LEARNER}
	\SetKwFunction{InitializeParam}{INITIALIZE\_PARAMETERS}
	\SetKwFunction{DDPGLoss}{COMPUTE_LOSS}
	\SetKwFunction{Apply}{ALLOCATE}
	\SetKwFunction{store}{STORE}
	\SetKwFunction{sample}{SAMPLE}
	\SetKwFunction{computeLoss}{COMPUTE\_DDPG\_LOSS}
	\SetKwFunction{updateParameters}{UPDATE\_PARAMETERS}

	\SetKwData{Learner}{central\_learner}
	\SetKwData{param}{param}
	\SetKwData{NFchains}{nfs}
	\SetKwData{Controller}{controller}
	\SetKwData{LocalBuffer}{local\_buffer}
	\SetKwData{ReplayBuffer}{replay\_buffer}
	
    \BlankLine
	
    \SetKwProg{procj}{Procedure}{}{\KwRet}
    \procj{\Actor{} }{
        \remotecall{\Learner.\param} \tcp{\scriptsize Collect latest update from the Central Learner}
        $x_i \leftarrow $ \getStat{\NFchains[1:N]} \tcp{\scriptsize collect the throughput, energy, packet arrival rate, CPU utilization}
        \For{ $i\gets1$ \KwTo $K$ } {
             $ a_{i} \leftarrow \pi_{i}(x_{i})$ \tcp{\scriptsize Get resource allocation from policy, $\pi$}  
             $[T_{i+1}, E_{i+1}, \xi_{i+1}, \Omega_{i+1}], r_{i+1}$ $\leftarrow$  \Controller.\Apply{$a_{i}$}  \\
             \LocalBuffer.\store{ $x_i, a_i, [T_{i+1}, E_{i+1}, \xi_{i+1}, \Omega_{i+1}], \gamma_{i+1}$ } \\
             Periodically :
             \ReplayBuffer.\store{\LocalBuffer} \\
             \remotecall{\Learner.\param} \tcp{\scriptsize Collect updates from central learner}
        }
    }
    \BlankLine
    \SetKwProg{proci}{Process}{}{\KwRet}
    \proci{\CentralLearner{} }{
        \InitializeParam{} \tcp{\scriptsize Initialize networks} 
        \For{ $i\gets1$ \KwTo $K$ } {
            \ReplayBuffer.\sample{} \tcp{ \scriptsize Prioritize experience sampling} 
            \computeLoss{}  \\
            \updateParameters{} \\
            \textrm{periodically remove the old experiences from replay buffer}
        }
     }   
    \caption{\name Framework}
\label{algo:GreenNFV}
\end{algorithm}

\textbf{Action Space: }
In \name, we define \textit{Action space} as the set of controllable resources in all NFs, $\mathcal{A}=\{\mathcal{A}_1, \mathcal{A}_2, ..., \mathcal{A}_n\}$ and for each NF the set of actions, $\mathcal{A}_i$ are - CPU core, $c_i$, CPU frequency, $cf_i$, last level cache, $llc_i$, DMA buffer size, $b_i$, batch size, $bs_i$. Therefore, for any NF the set of actions are, 
\begin{equation}
    \mathcal{A}_i = \{c_i, cf_i, llc_i, b_i, bs_i\}
\end{equation}

\textbf{State Space:}
The state space is defined as the set of current status for all NF as, $\mathcal{X} = \{ \mathcal{X}_1,\mathcal{X}_2, ..., \mathcal{X}_n\}$ that contains throughput, $T_i$, energy consumption, $E_i$, CPU utilization, $\xi_i$, packet arrival rate, $\Omega_i$. Therefore, $\forall \mathcal{X}_i \in \mathcal{X}$, we have:
\begin{equation}
    \mathcal{X}_i = \{T_i, E_i, \xi_i, \Omega_i\}
\end{equation}

\textbf{Reward Signal: }
In \name, we have three different reward functions for different SLA-based resource optimization that are elaborated in $\S$\ref{subsec:design_considerations}. For each SLA, we define different reward signals. The reward is the response from the environment to the learning agent to measure how good its current actions are from the previous step. 
To maximize the throughput SLA, we define the reward function as described in Equation~\ref{eq:throughput_sla}.
For minimizing the energy SLA, we define $\psi_{E}$ (Equation~\ref{eq:energy_sla}) as the reward function. 
For the energy efficiency SLA, we defined the reward function as explained in Equation~\ref{eq:energy_efficiency_sla}.

\subsubsection{Distributed Learning Framework}

On top of DDPG, \name utilizes Ape-X framework~\cite{horgan2018distributed}. The ape-X framework uses a distributed architecture to scale up the DDPG algorithm. It extends the concept of prioritized experience replay to achieve the goal. 
Experience replay~\cite{lin1992self} has been heavily used in reinforcement learning to improve data efficiency. It becomes highly effective to train DNNs~\cite{mnih2015human} as it utilizes a buffer of previous experiences to stabilize the training procedure. 
The experience samples can be defined as a tuple in the form of $(x_i,a_i,r_i,x_{i+1})$ with states, actions, rewards, and successor states at time t. 

In contrast to consuming samples online and discarding them later, sampling from the stored experiences means they are less heavily “correlated” and can be reused for learning. 
Experience replay prevents the overfitting of the model by allowing the agent to learn from previous policy versions. The procedure is closely related to importance sampling.
Uniform sampling from a replay buffer is a good default strategy and probably the first to attempt. However, prioritized experience sampling~\cite{Schaul2016}, as the name implies, will weigh the samples so that “important” ones are drawn more frequently for training.

In this setup, the actor and learner modules can be distributed across multiple workers. Actors run on servers and generate data according to the current policy. A single learner samples the new experience and updates the policy parameters. These updated parameters are sent periodically to the actors.
This framework implements a centralized replay memory with prioritized experience replay. 
%
The framework is explained in Algorithm~\ref{algo:GreenNFV} that explains the steps of  \texttt{NF\_CONTROLLER} (actor) and a central learner process. Initially, \texttt{NF\_CONTROLLER} performs a remote call to the learner process and receives the current parameters to update its policy. Then the actor collects information about the current network function chains and the flows. Based on this information, \texttt{NF\_CONTROLLER} computes the current resource control settings (actions) using its current policy and reconfigures the resources accordingly. Then the controller stores this information as a sample experience in its local buffer. Periodically, the \texttt{NF\_CONTROLLER} sends the content of the local buffer to the central replay buffer and collects the latest parameter values for the policy.
The \texttt{CENTRAL\_LEARNER} process periodically collects prioritized experiences from the replay buffer. Then it computes the DDPG loss function and updates the parameters. Periodically, it removes old experiences from the replay buffer.

%% file: section-Implementation.tex
\begin{figure}[t]
\centering
\includegraphics[keepaspectratio=true,angle=0,width=80mm]{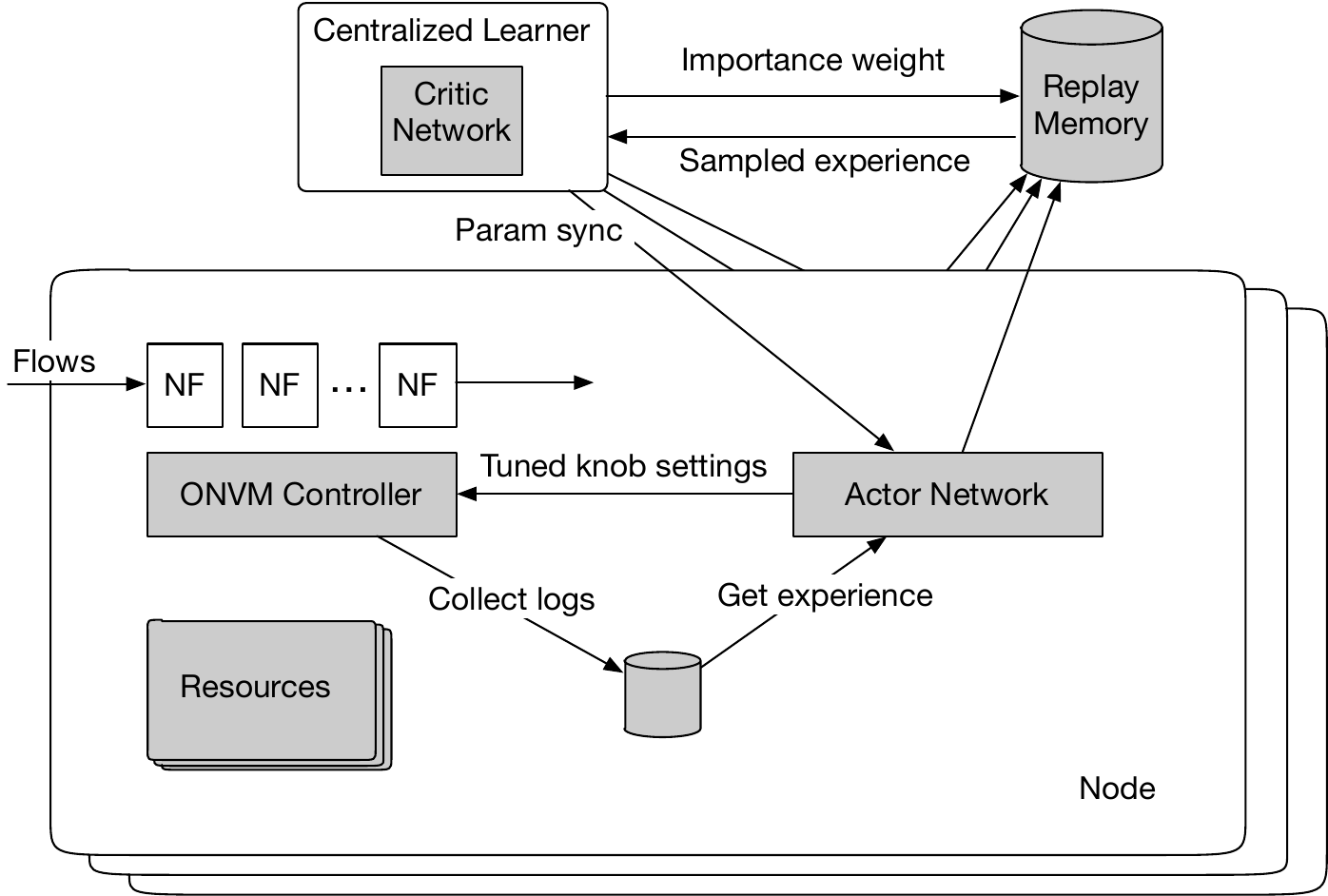}
\caption{Overview of the \name architecture.} \label{fig:model_overview}
\end{figure}

\subsection{\name Implementation}
\label{sec:implementation}

\begin{figure*}[t]
    \begin{centering}
\begin{subfigure}[t]{0.24\textwidth}
    \centering
        \includegraphics[keepaspectratio=true,width=40mm]{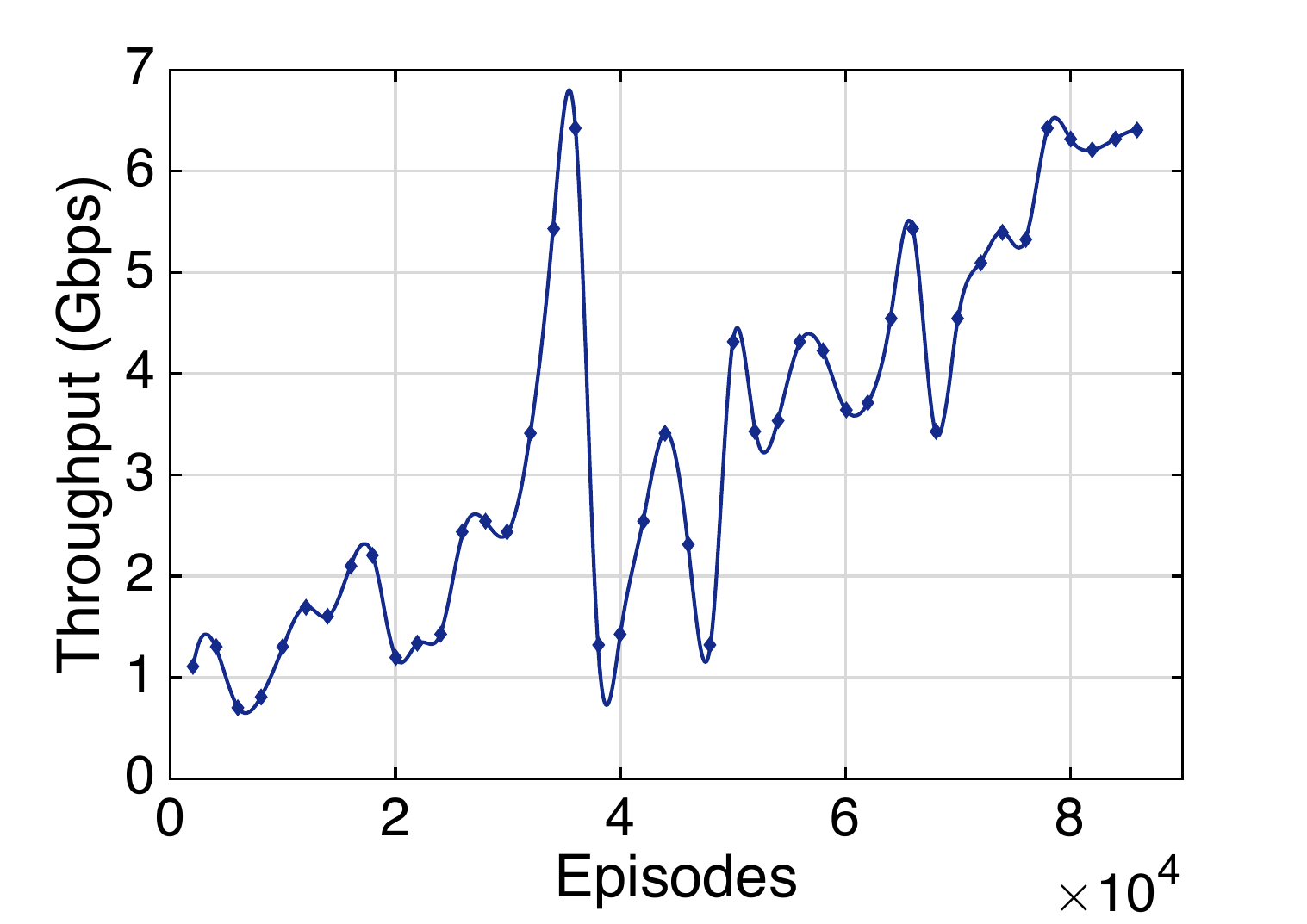}
        \caption{Achieved throughput}
\end{subfigure}
\begin{subfigure}[t]{.24\textwidth}
        \includegraphics[keepaspectratio=true,width=40mm]{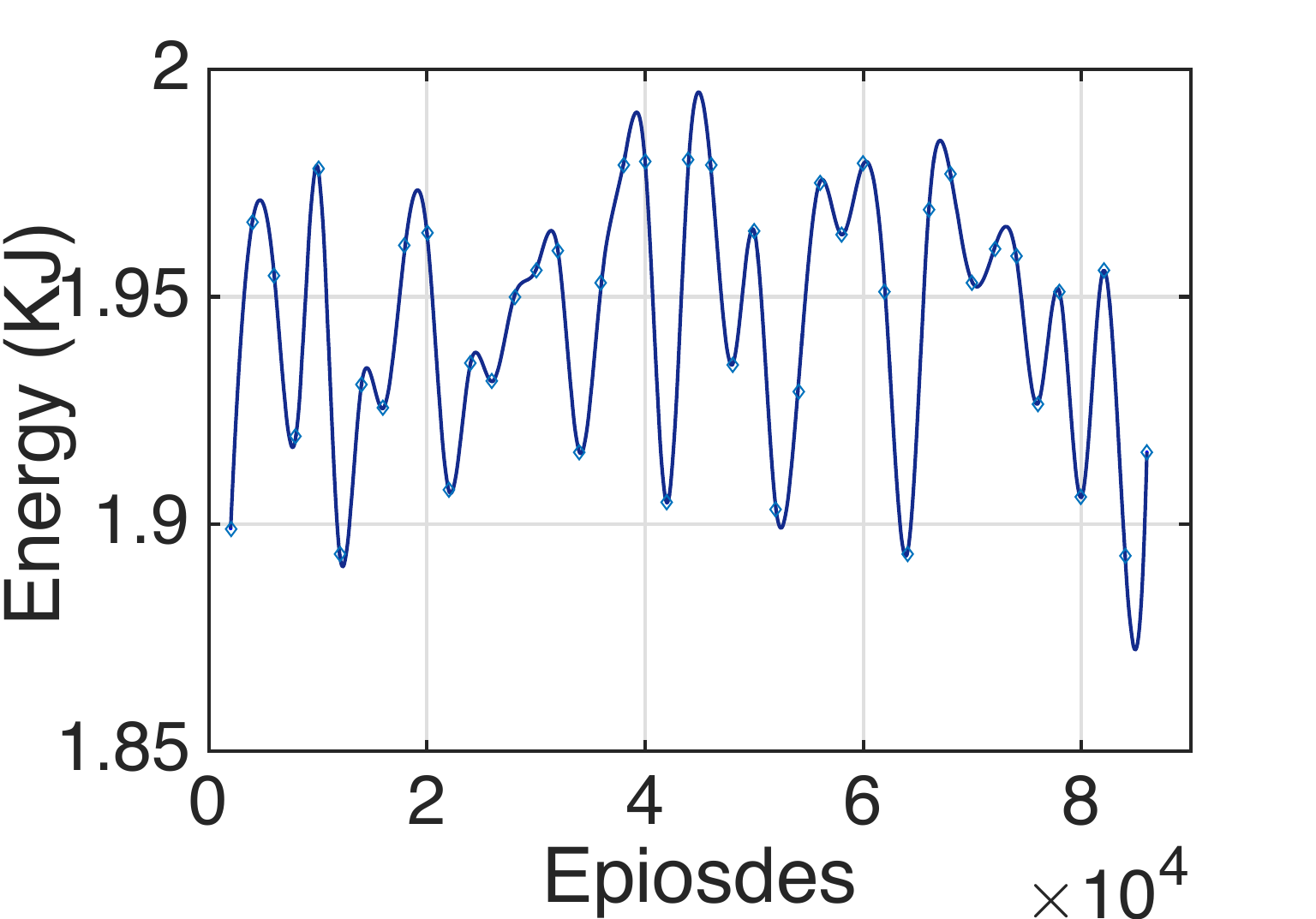}
        \caption{Energy consumption}
\end{subfigure}
\begin{subfigure}[t]{0.24\textwidth}
        \includegraphics[keepaspectratio=true,width=40mm]{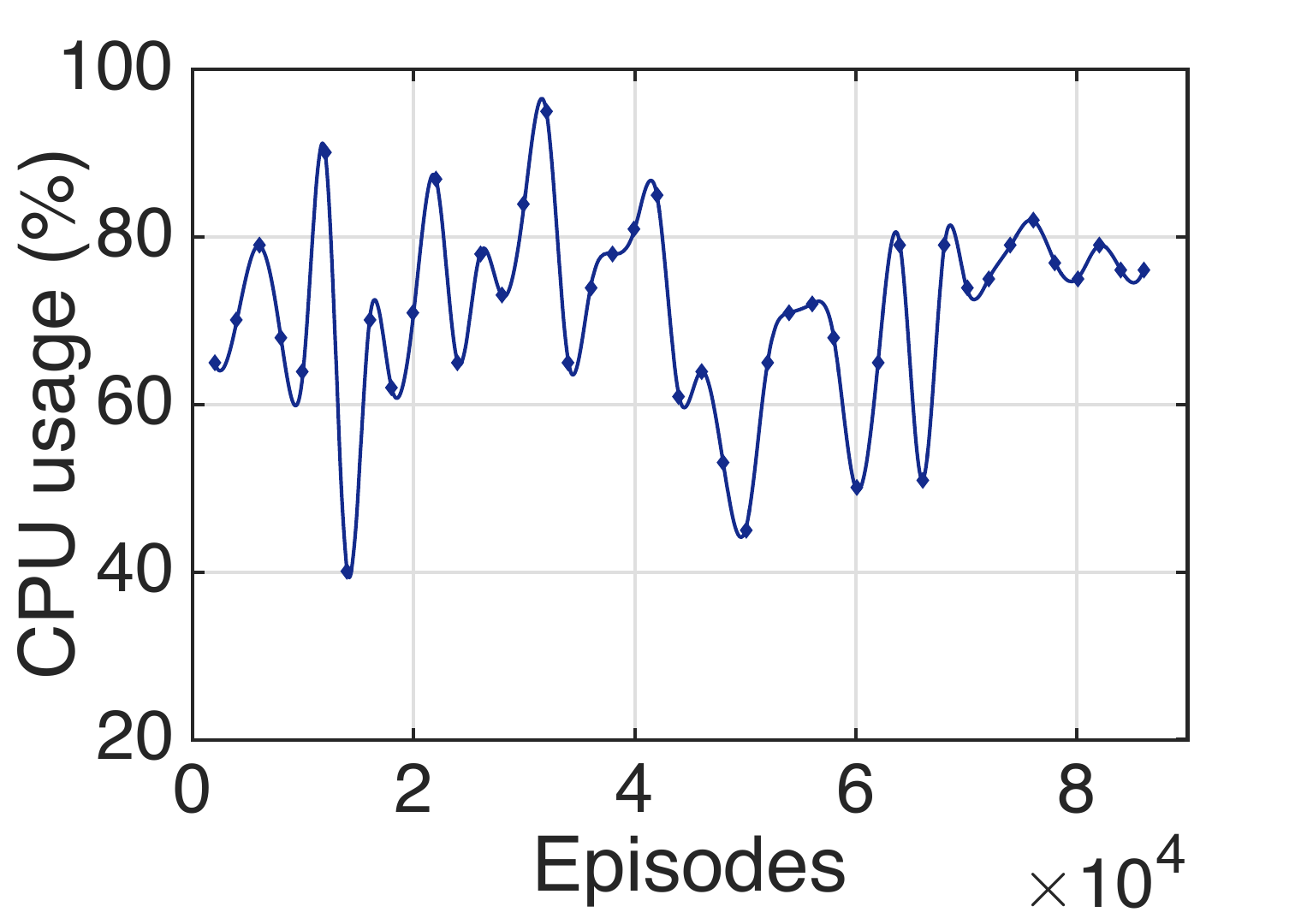}
        \caption{CPU utilization}
\end{subfigure}
\begin{subfigure}[t]{0.24\textwidth}
        \includegraphics[keepaspectratio=true,width=40mm]{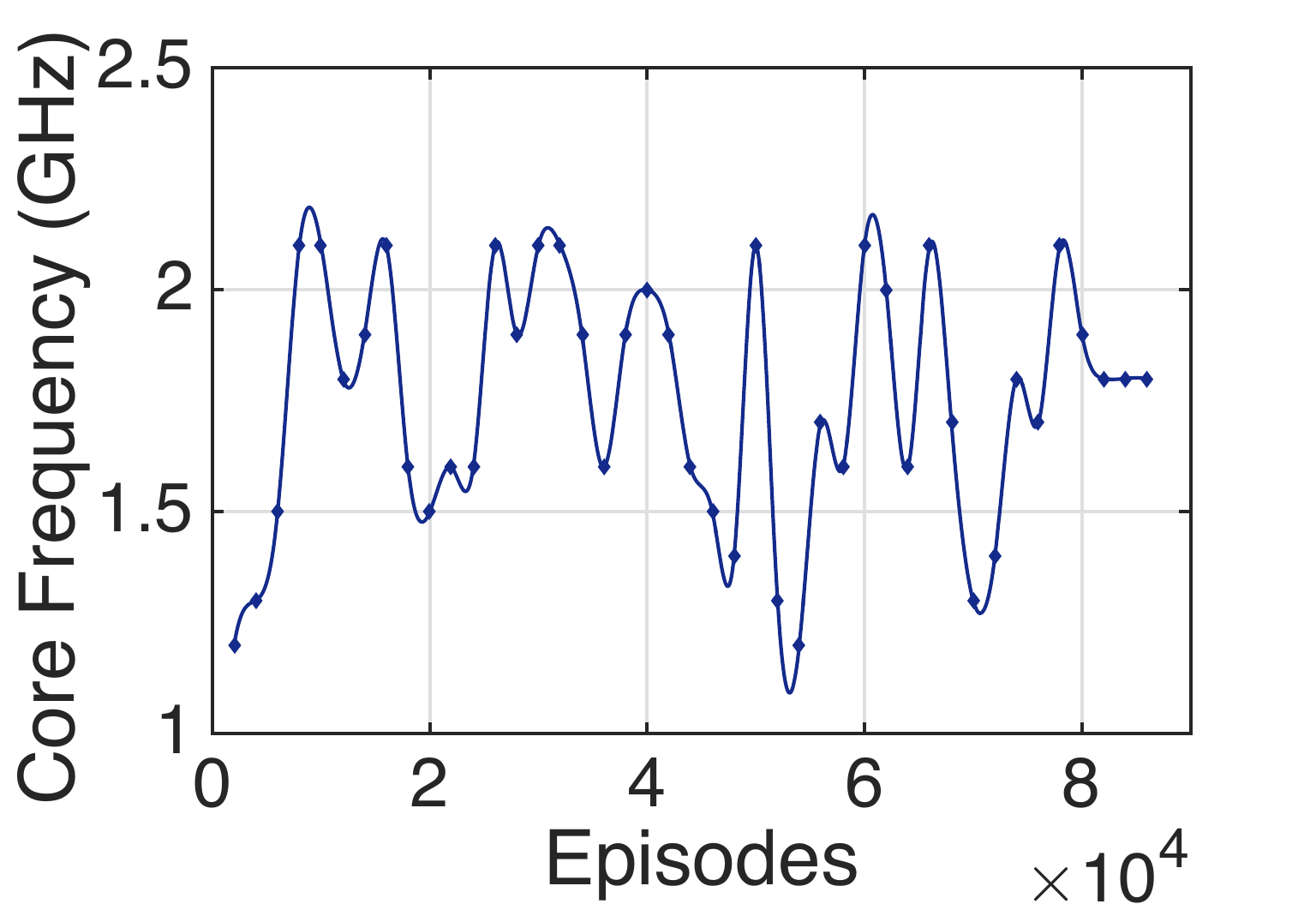}
        \caption{CPU Core Frequency}
\end{subfigure}
\begin{subfigure}[t]{0.24\textwidth}
        \includegraphics[keepaspectratio=true,width=40mm]{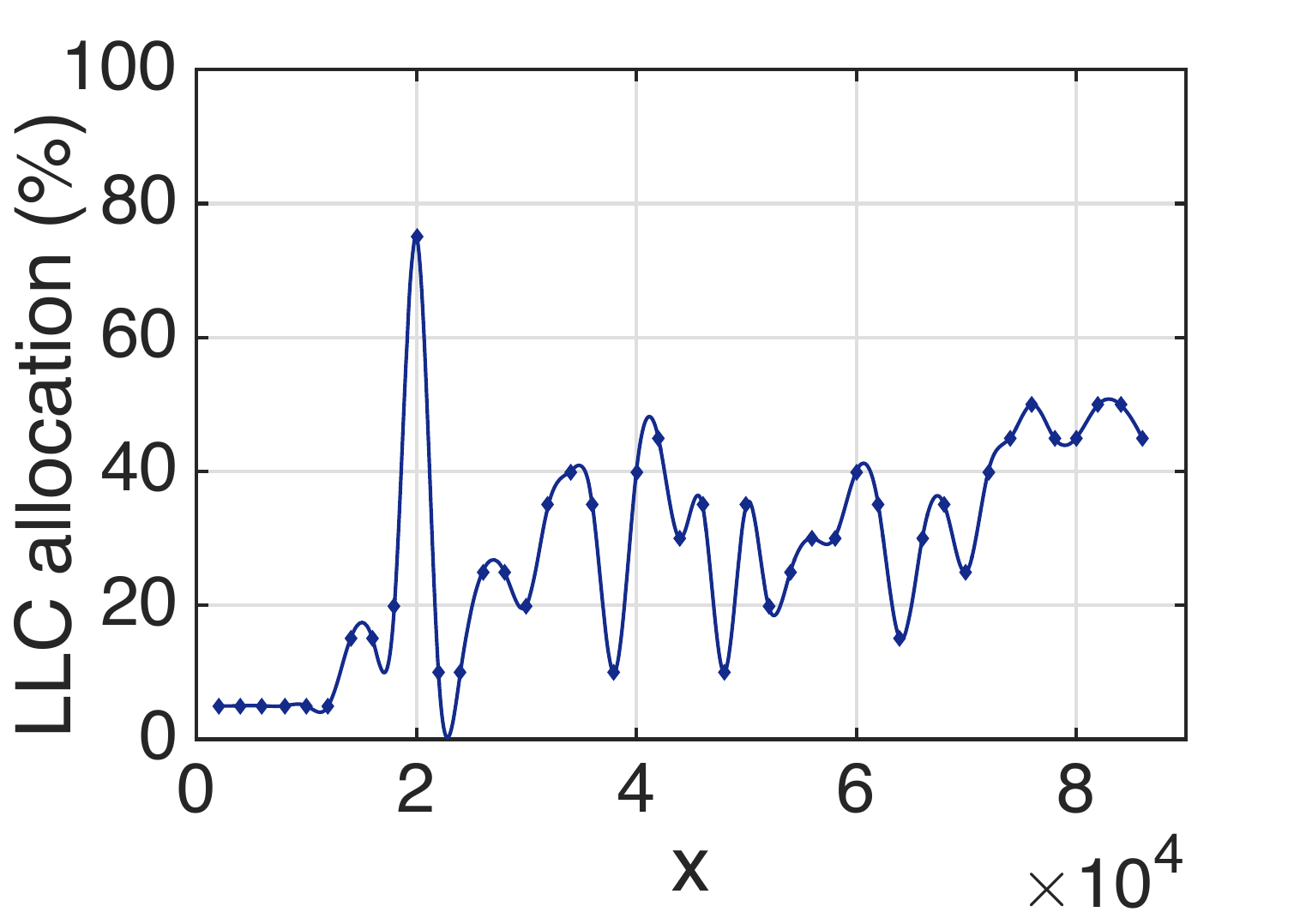}
        \subcaption{LLC allocation}
\end{subfigure}
\begin{subfigure}[t]{0.24\textwidth}
        \includegraphics[keepaspectratio=true,width=40mm]{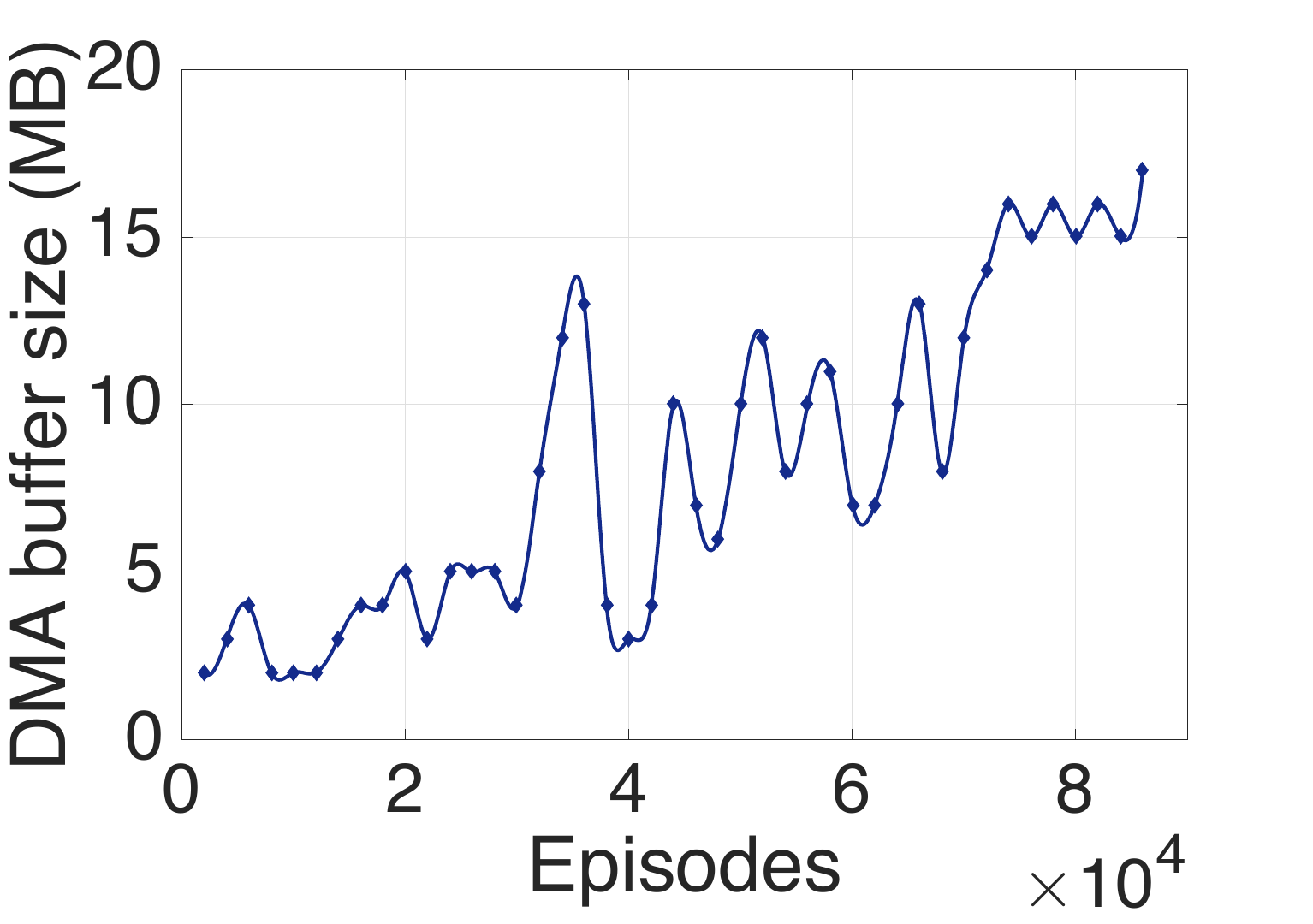}
        \caption{DMA buffer size}
\end{subfigure}
\begin{subfigure}[t]{0.24\textwidth}
        \includegraphics[keepaspectratio=true,width=40mm]{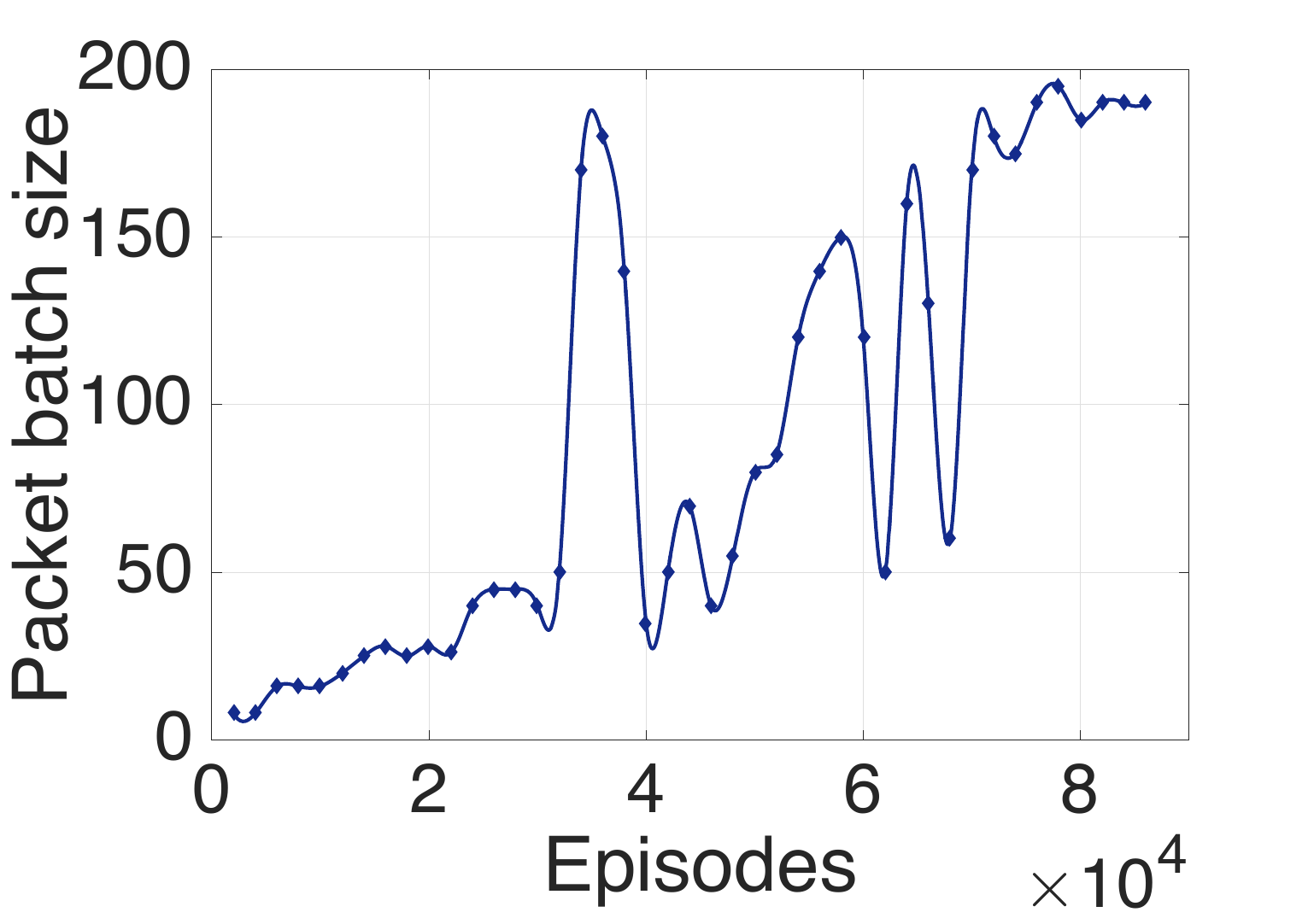}
        \caption{Packet batch size}
\end{subfigure}
     \caption{Training progress of the proposed reinforcement learning algorithm during the testing of the  Maximum Throughput SLA.}
     \label{fig:SLA_maxth_results}
     \end{centering}
 \end{figure*}


\name is implemented over the OpenNetVM~\cite{zhang_opennetvm:_2016}, a popular packet processing framework to develop, deploy and manage network functions. It uses the Intel DPDK library for high-speed packet processing. The platform runs NFs as individual process or in a docker container. Each NF has two circular queues to track incoming and outgoing packets. OpenNetVM (ONVM) controller has Rx and Tx threads, running on a dedicated core, to manage packet circulation through NFs. NFs can run on an individual core or multiple cores or even share the same core depending on the processing requirements. Some NFs work in a distributed manner, thus requiring coordination. Multiple IDS VNFs can be deployed and share information for more fine-grained protection. 

In \name implementation, one of the main challenges is to design NF management in a way that can dynamically conserve energy while maintaining performance. When the packet arrival rate increases, we want to allocate more resources to the NF. However, when the packet arrival rate slows down, the CPU should go to a low power state, and batch size and LLC allocation should change too. When there is no packet to process, we put NF to sleep until a new packet arrives. 

Figure~\ref{fig:model_overview} shows typical NF chains deployed in multiple nodes. However, such an assignment requires careful consideration based on the compute and memory requirement of NFs, and packet arrival rates. Service chains can be configured using a configuration file or SDN controller. 
We added functionalities in the ONVM controller that allow us to control the CPU share, DVFS (CPU frequency) control, LLC allocation, DMA Buffer size, and packet batch size.
To access the DVFS, we use \texttt{cpufrequtils}~\cite{cpufrequtil} library. It can provide fine-grained access to the CPU frequencies. The library provides numerous power management schemes called power governors, such as userspace, on-demand, conservative, power-save, and performance. The Userspace governor enables frequency control from the userspace. We use  \texttt{userspace} governor to control the CPU frequencies of the individual CPU cores. 

To control the LLC allocation, we use Intel CAT~\cite{IntelCAT}. This Intel library only works on Intel processors, and it provides fine userspace control of the LLC. It is particularly helpful when multiple workflows share the same resources. This library provides access to a construct named Class of Services (CLOS) that can be used to group the NFs. It is possible to dynamically assign the LLC to these CLOS from the userspace. They also provide resource capacity bitmask (CBM) to allocate cache to each group.  
We implement CPU scheduling to the NF similar to NFVNice~\cite{Kulkarni:2017}. Linux provides a kernel feature, named \texttt{cgroups} that allows more fine-grained resource allocation - such as CPU time, system memory, network bandwidth, and their combinations. \texttt{cgroups} uses a similar hierarchy and inheritance like the processes.  
We also integrated a Distributed RL framework that can cooperate with the ONVM controller. The Actor network is responsible for computing resource allocation. ONVM controller can request resource allocation from the Actor-network. We also implemented a centralized learner and replay buffer, that periodically updates Actor networks with new parameters. The learning module is developed with Python 3.6 and Tensorflow~\cite{tensorflow} framework.

%% file: section-evaluation.tex
\section{Evaluation}
\label{sec:evaluation}

We evaluate \name under three different optimization goals based on differing Service Level Agreements (SLA): 

\begin{itemize}

\item Maximize the achievable throughput without violating the power constraint set by the SLA ($\S$\ref{subsec:throughput}). 

\item Minimize the energy consumption while providing a specific throughput guarantee
($\S$\ref{subsec:energy}). 

\item Maximize energy efficiency in terms of the throughput achieved per unit energy consumption ($\S$\ref{subsec:energy-efficiency}).
\end{itemize}

\name is evaluated using six nodes from a cloud service provider. Nodes are equipped with Intel Xeon CPU E5-2620 (v4) with frequencies ranging from 1.2 GHz to 2.1 GHz. Each CPU has dual sockets with a total of 16 cores. The main memory size for each server is 64 GB. The nodes are connected with a DPDK-compatible NIC card - Intel 10 Gigabit X540-AT2. Intel DPDK is a highly popular packet processing framework and is extensively used in Telecommunication systems. Hardware preferences can be generalized to any Processor or NIC that supports the DPDK framework. The servers are running Ubuntu SMP with Linux kernel 4.4.0-177-generic as the operating system. 
As the traffic generator software, we use MoonGen~\cite{moongen2015} that can generate UDP and TCP traffic at line rate. Packet size varies from 64 Byte to 1518 Byte.  In our experiment, we used three servers to generate the traffic using MoonGen, and the rest of the three servers are used to host the NF chains. Each node hosts an NF chain with three Network functions. Network functions are chained with a series connection.

We compare our model with the Energy Efficient P-state (EE-Pstate) approach from~\cite{faisal12}. In that work, the authors use a threshold-based approach to decide on P-state. They also use simple predictors like - Double Exponent Smoothing Predictor (DES) for traffic prediction. We also compare our model with the baseline model that uses a Performance power governor, and all other components are set to default values. Finally, we compare our model with the Heuristics model ($\S$\ref{subsec:heuristic}) and the Q-learning model. For the Q-learning model, we discretize the action and state space. 

\subsection{Maximum Throughput SLA}
\label{subsec:throughput}

In this experiment, we evaluate the Maximum Throughput SLA, where the model aims to maintain the best achievable throughput without violating the SLA's energy constraint. We set the maximum energy threshold to $2000$ Joules and use five flows. During the training process, we test the performance periodically at each 2000th episode. The reward function used in this SLA issues rewards only when the agent can meet the energy SLA.

\begin{figure*}[t]
    \begin{centering}
\begin{subfigure}[t]{0.24\textwidth}
    \centering
        \includegraphics[keepaspectratio=true,width=40mm]{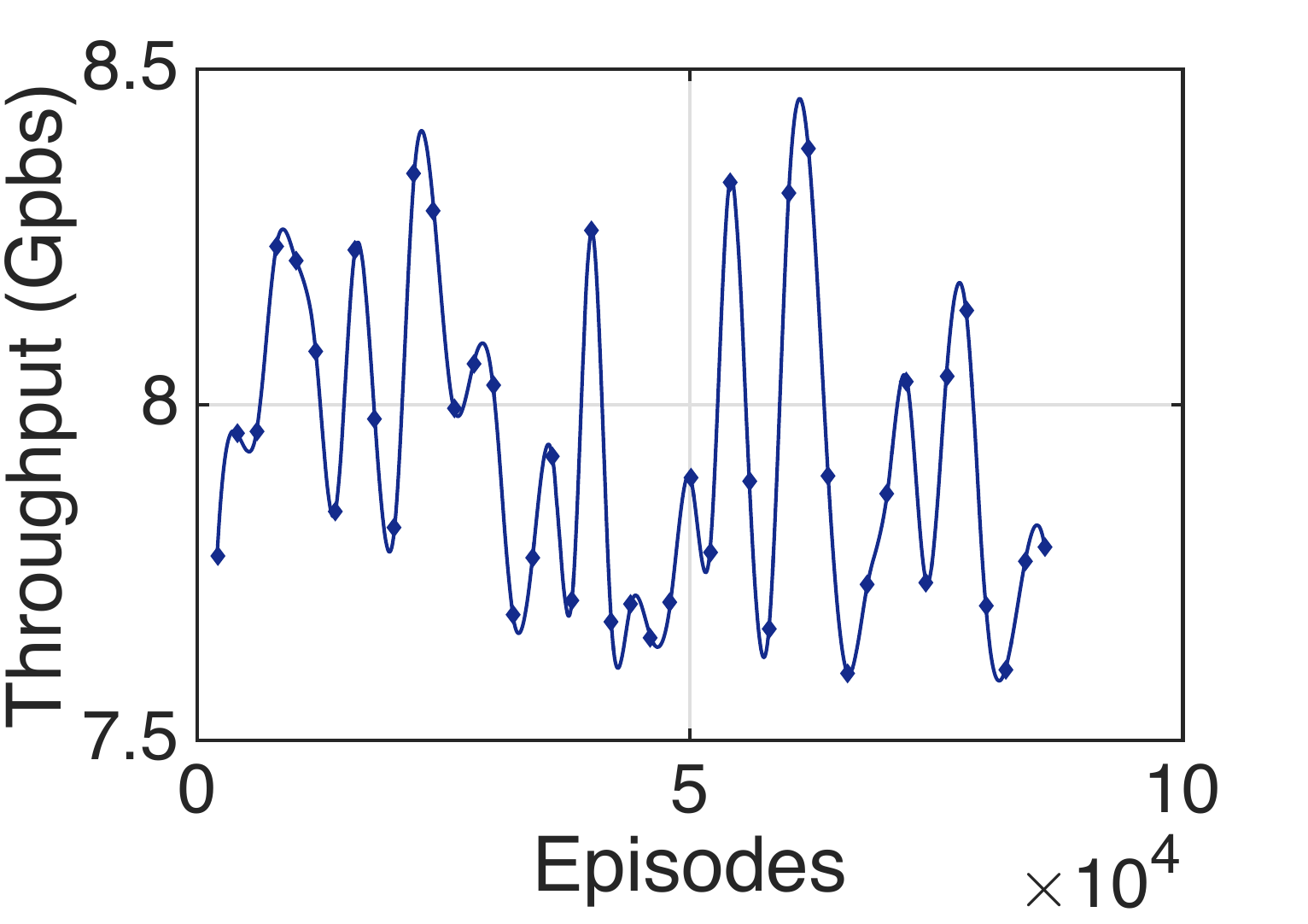}
        \caption{Achieved throughput}
\end{subfigure}
\begin{subfigure}[t]{.24\textwidth}
        \includegraphics[keepaspectratio=true,width=40mm]{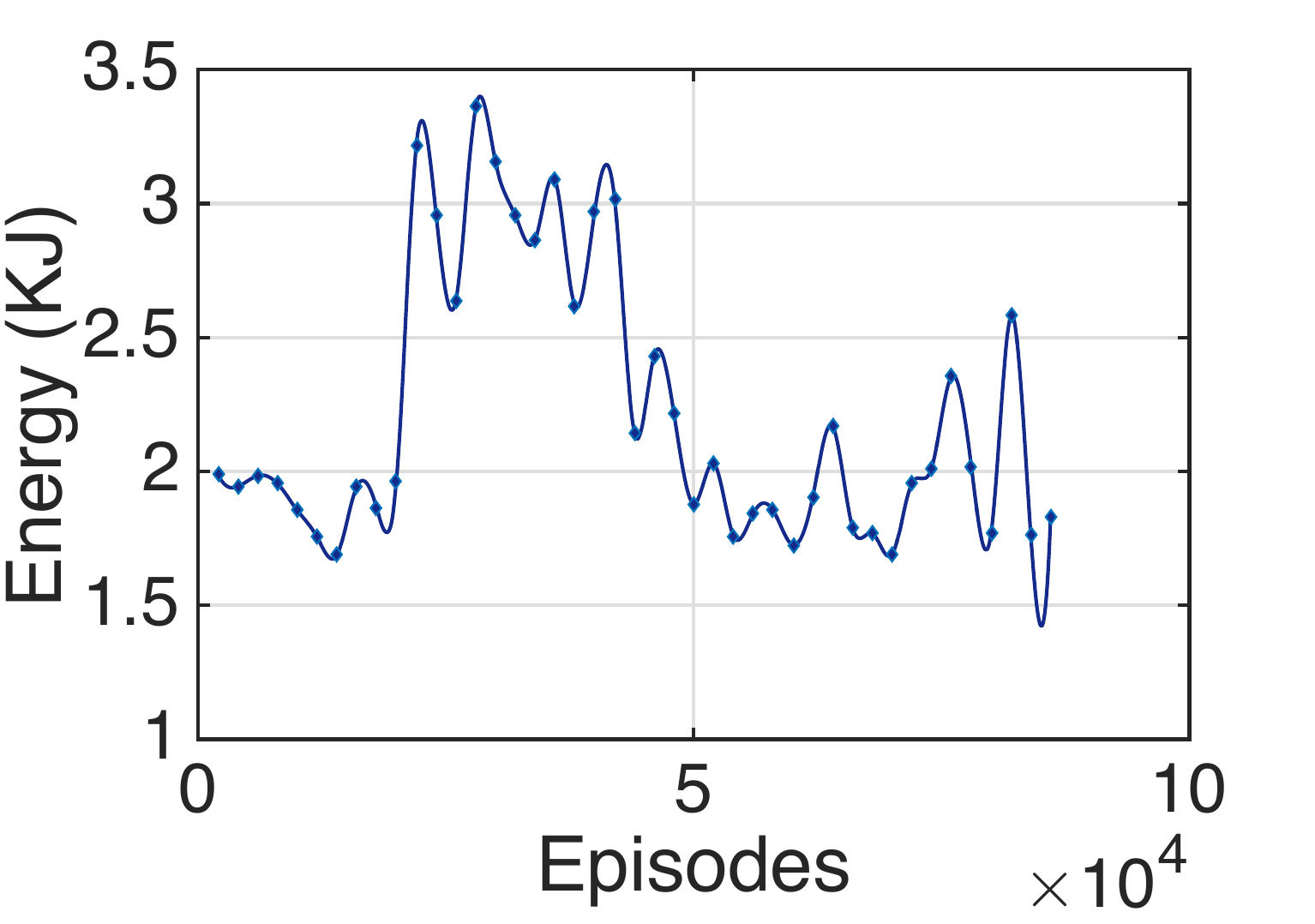}
        \caption{Energy consumption}
\end{subfigure}
\begin{subfigure}[t]{0.24\textwidth}
        \includegraphics[keepaspectratio=true,width=40mm]{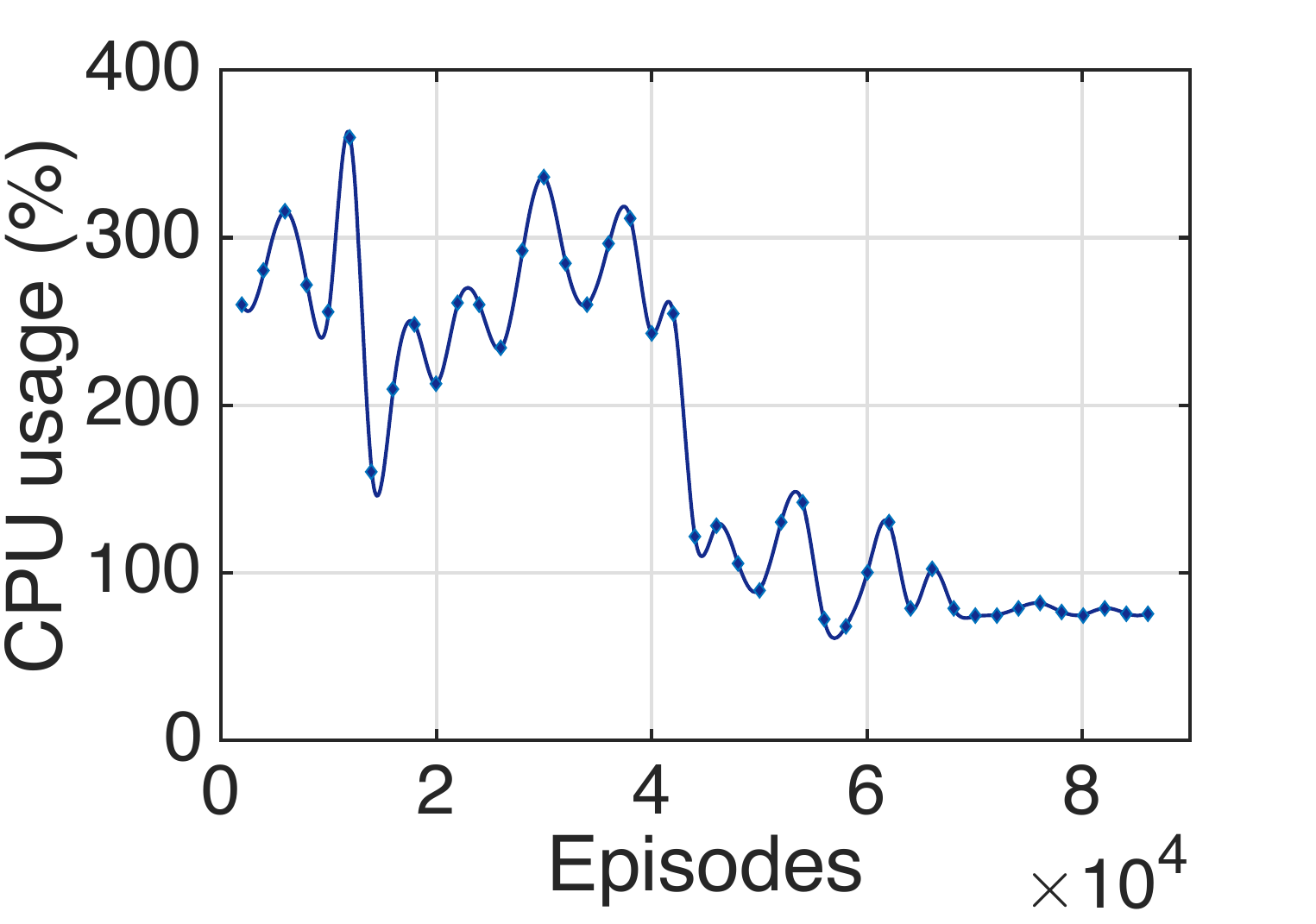}
        \caption{CPU utilization}
\end{subfigure}
\begin{subfigure}[t]{0.24\textwidth}
        \includegraphics[keepaspectratio=true,width=40mm]{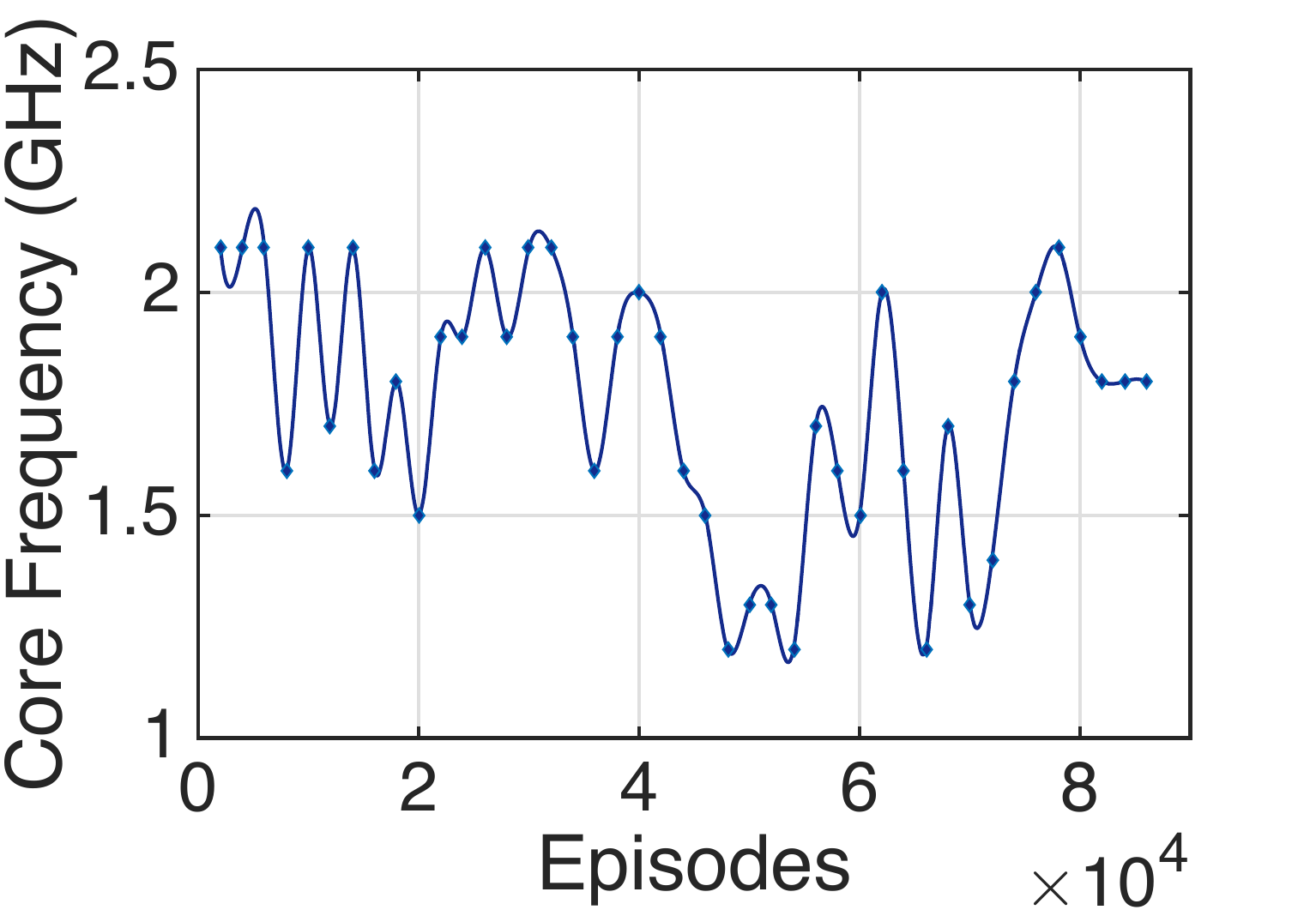}
        \caption{CPU core frequency}
\end{subfigure}
\begin{subfigure}[t]{0.24\textwidth}
        \includegraphics[keepaspectratio=true,width=40mm]{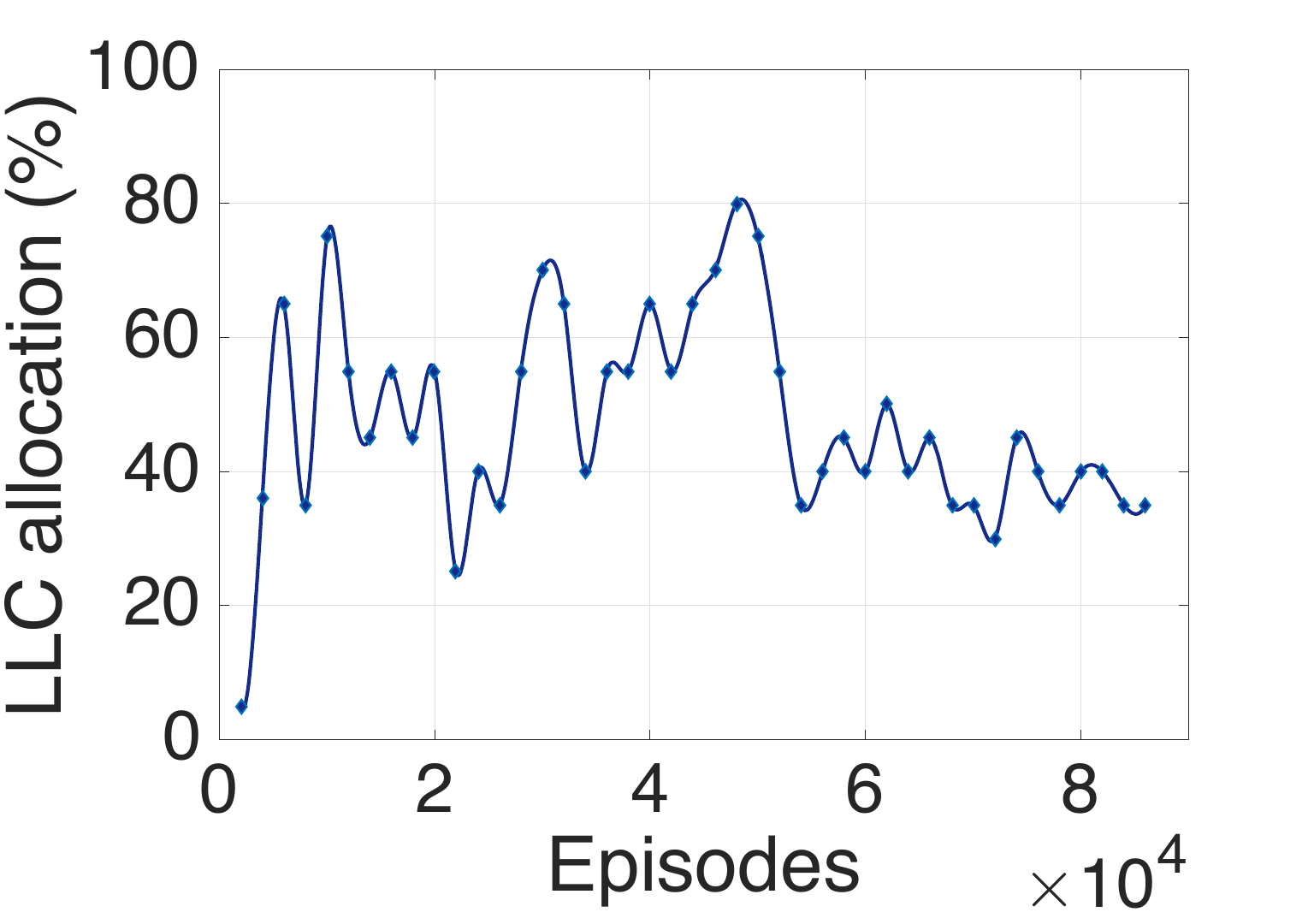}
        \subcaption{LLC allocation}
\end{subfigure}
\begin{subfigure}[t]{0.24\textwidth}
        \includegraphics[keepaspectratio=true,width=40mm]{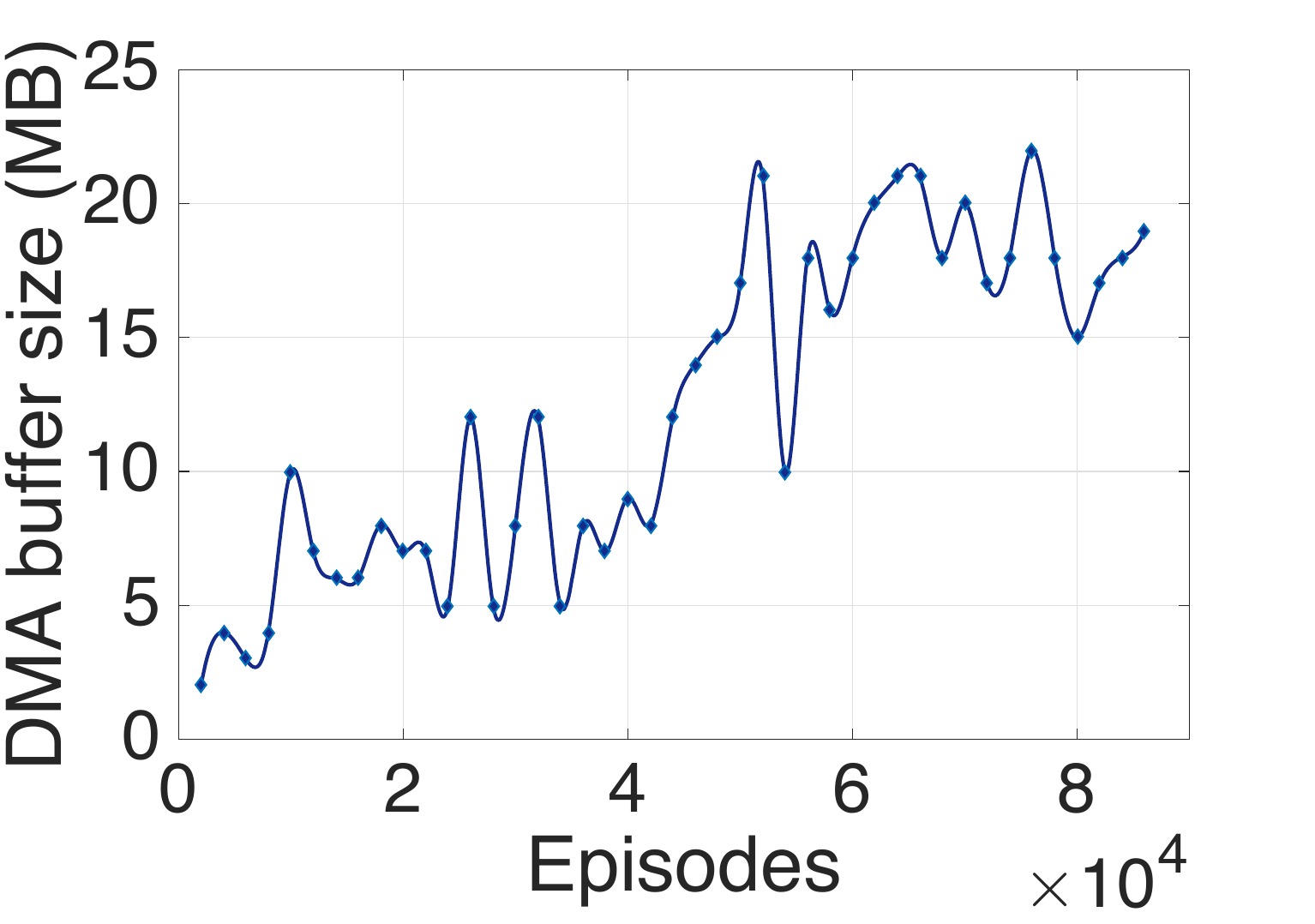}
        \caption{DMA buffer size}
\end{subfigure}
\begin{subfigure}[t]{0.24\textwidth}
        \includegraphics[keepaspectratio=true,width=40mm]{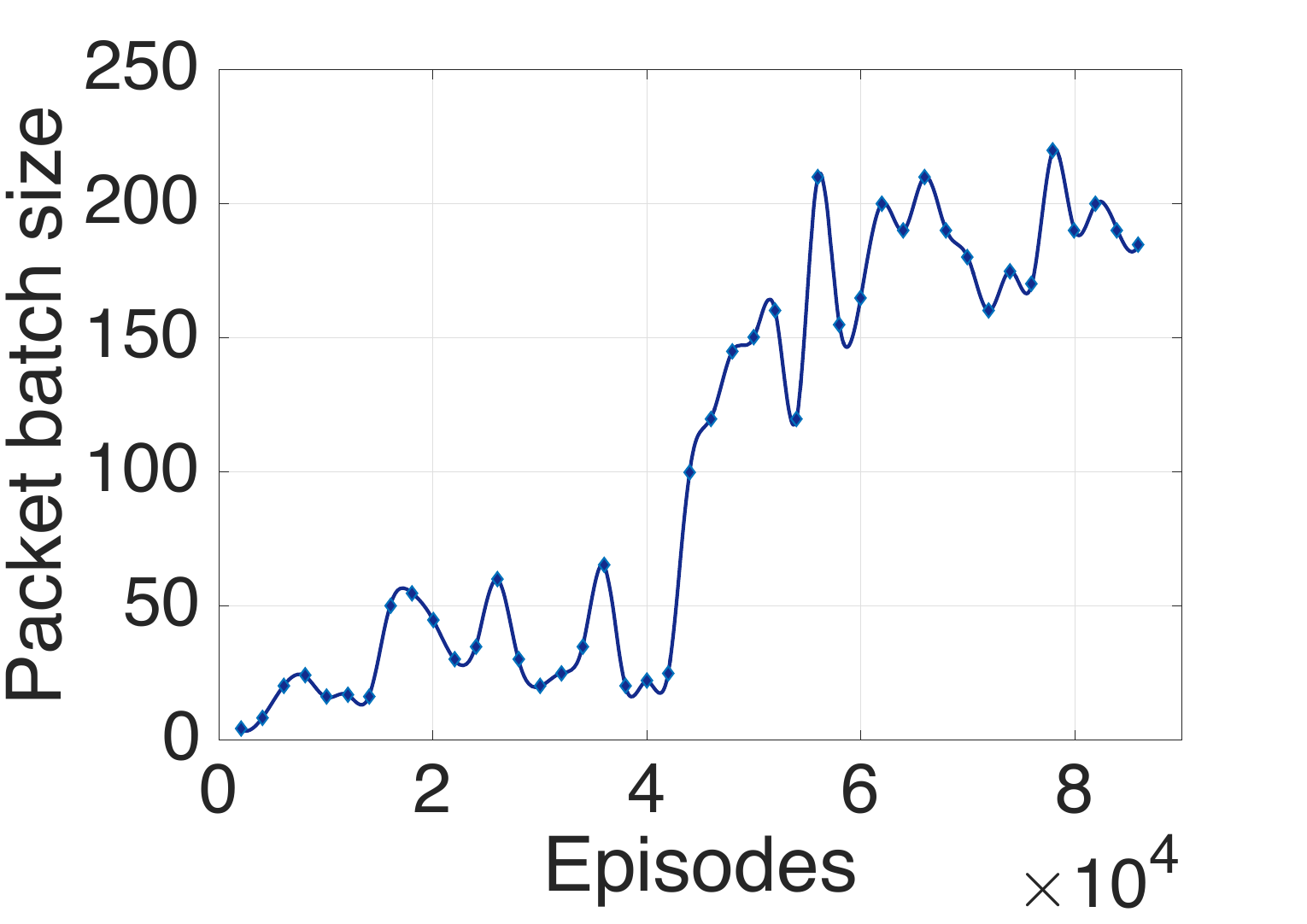}
        \caption{Packet batch size}
\end{subfigure}
     \caption{Training progress of the proposed reinforcement learning algorithm during the testing of the  Minimum Energy SLA.}
     \label{fig:SLA_minE_results}
     \end{centering}
 \end{figure*}

Figures~\ref{fig:SLA_maxth_results}(a-b) show the throughput and energy consumption as the training progresses. The throughput value increases during the training process (Figure~\ref{fig:SLA_maxth_results}(a)). 
As shown in Figure~\ref{fig:SLA_maxth_results}(b), the model learns to restrict the energy consumption below the energy constraint set by the SLA. Then it tries to increase the throughput during the learning process.
Figures~\ref{fig:SLA_maxth_results}(c-g) shows the selection of each control component. When the batch size of the packets is increased, it helps to achieve high throughput while maintaining the energy consumption below a certain threshold (Figure~\ref{fig:SLA_maxth_results}(c)). The batch size is a critical component to achieve high performance, as it reduces the cache miss rate that can arise due to packet loading from the main memory. Passing one packet at a time requires the function call for each packet. However, packets in batches reduce the number of function calls drastically and increase the performance without a high energy toll. 
The LLC allocation is also increased to achieve higher performance so that the batch can be processed faster without many cache misses (Figure~\ref{fig:SLA_maxth_results}(d)). Increased DMA buffer size also helps with the increase in the achieved throughput (Figure~\ref{fig:SLA_maxth_results}(e)). The model performs a balance to accommodate the batch of packets by controlling the buffer size. These three control knobs (packet batch size, LLC allocation, and DMA buffer size) are directly contributing to the increase in the achievable throughput and have less impact on energy consumption.

\begin{figure*}[t]
    \begin{centering}
\begin{subfigure}[t]{0.24\textwidth}
    \centering
        \includegraphics[keepaspectratio=true,width=40mm]{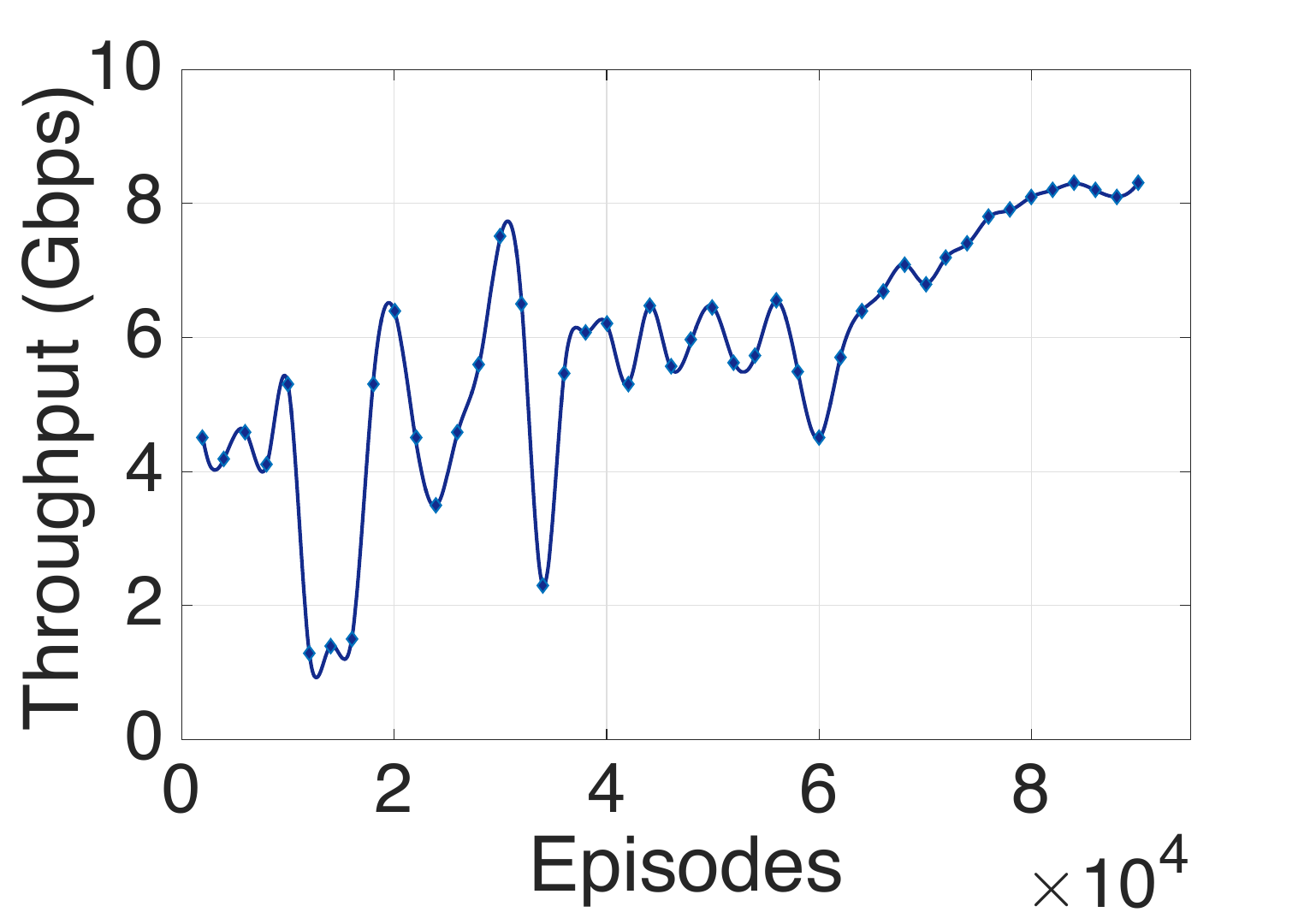}
        \caption{Achieved throughput}
\end{subfigure}
\begin{subfigure}[t]{.24\textwidth}
        \includegraphics[keepaspectratio=true,width=40mm]{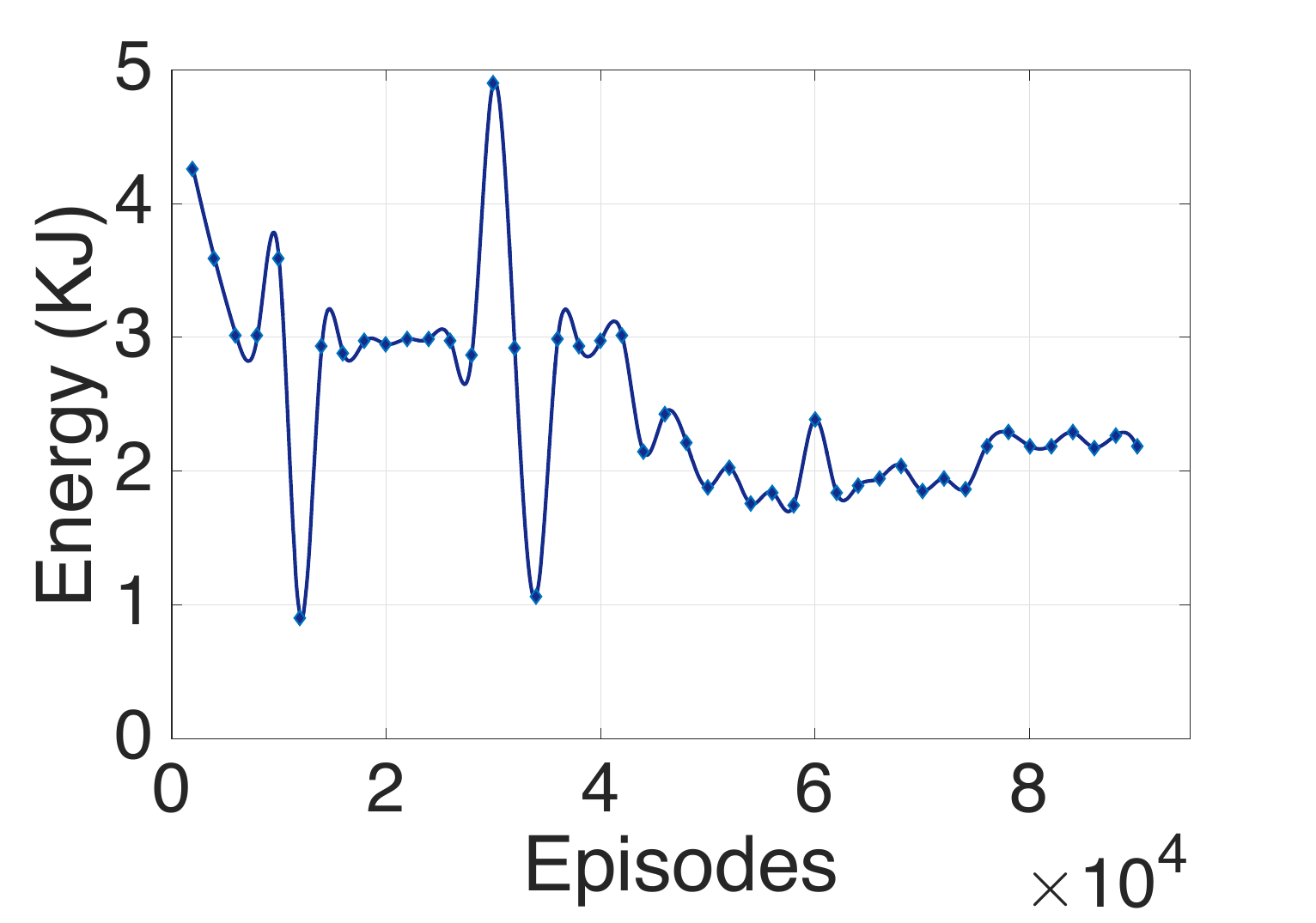}
        \caption{Energy consumption}
\end{subfigure}
\begin{subfigure}[t]{0.24\textwidth}
        \includegraphics[keepaspectratio=true,width=40mm]{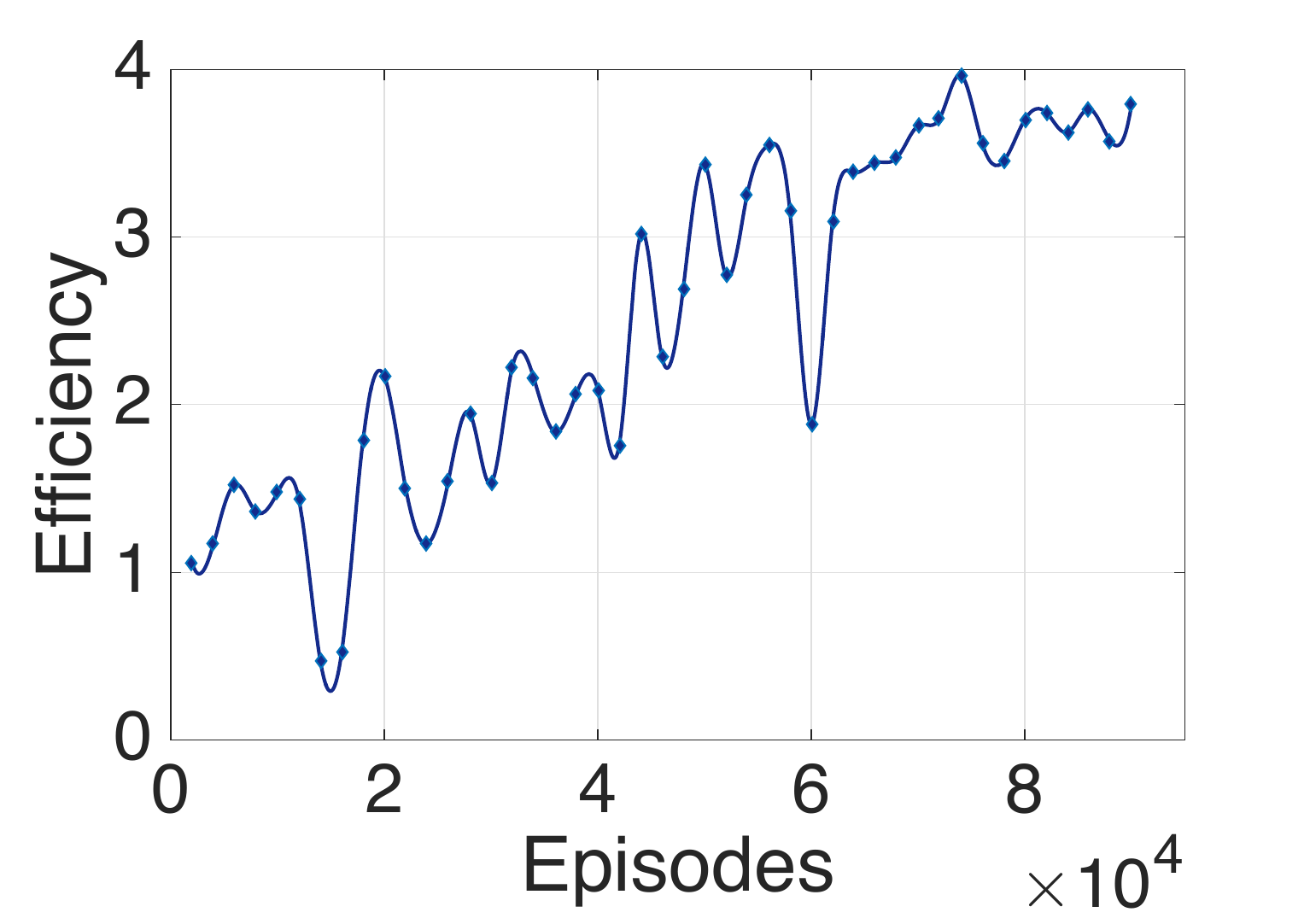}
        \caption{Energy Efficiency}
\end{subfigure}
\begin{subfigure}[t]{0.24\textwidth}
        \includegraphics[keepaspectratio=true,width=40mm]{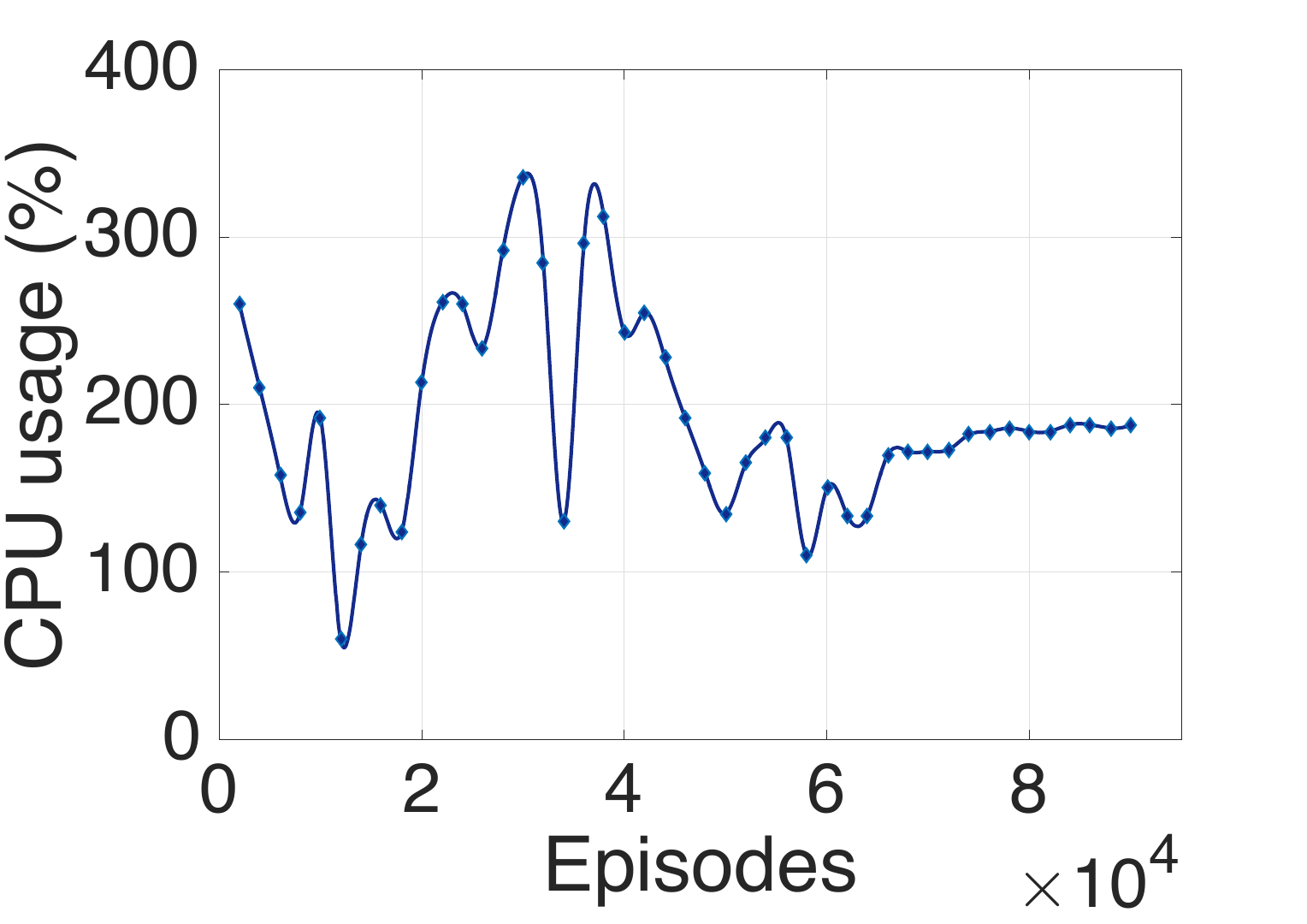}
        \caption{CPU utilization}
\end{subfigure}
\begin{subfigure}[t]{0.24\textwidth}
        \includegraphics[keepaspectratio=true,width=40mm]{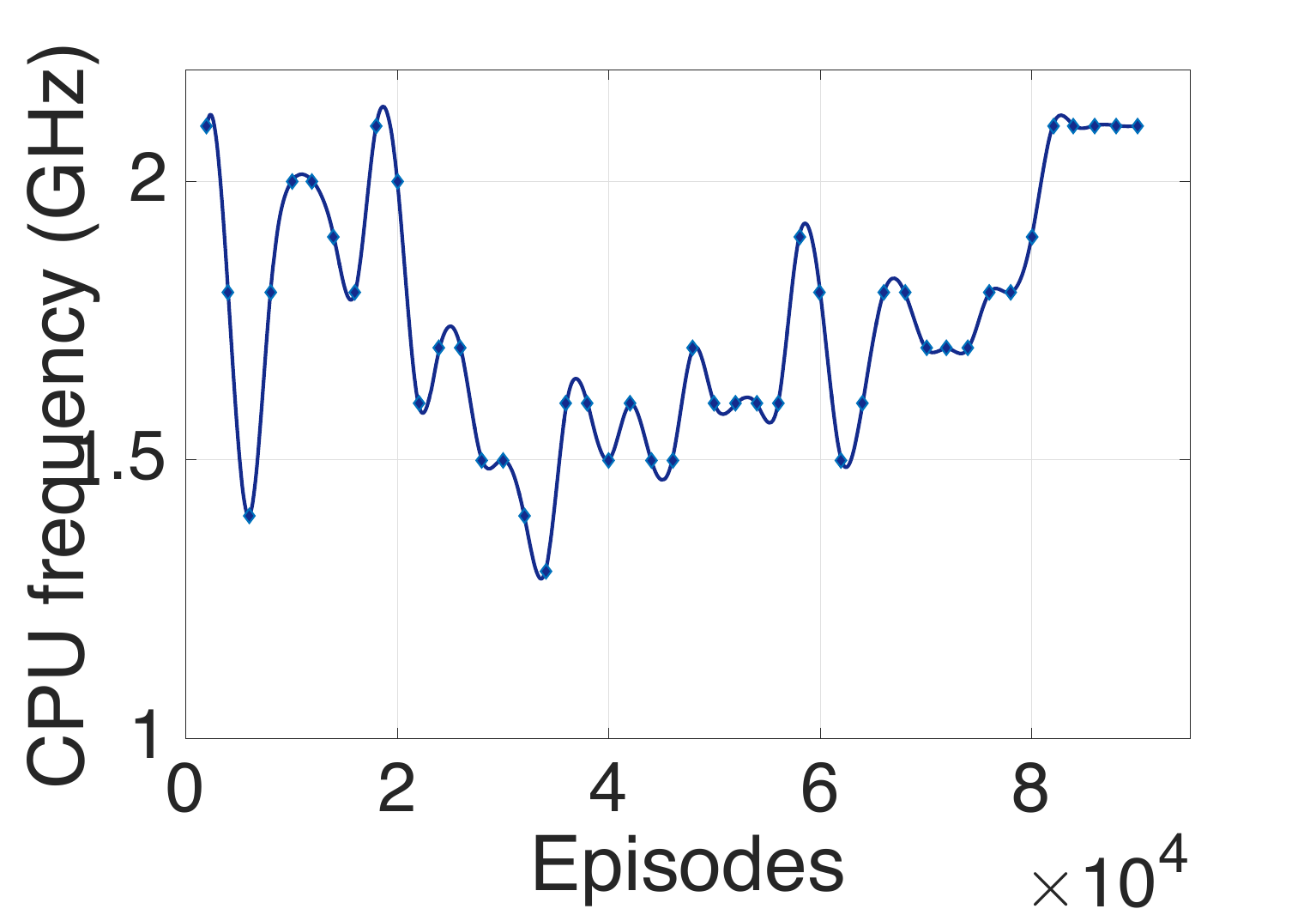}
        \caption{CPU core frequency}
\end{subfigure}
\begin{subfigure}[t]{0.24\textwidth}
        \includegraphics[keepaspectratio=true,width=40mm]{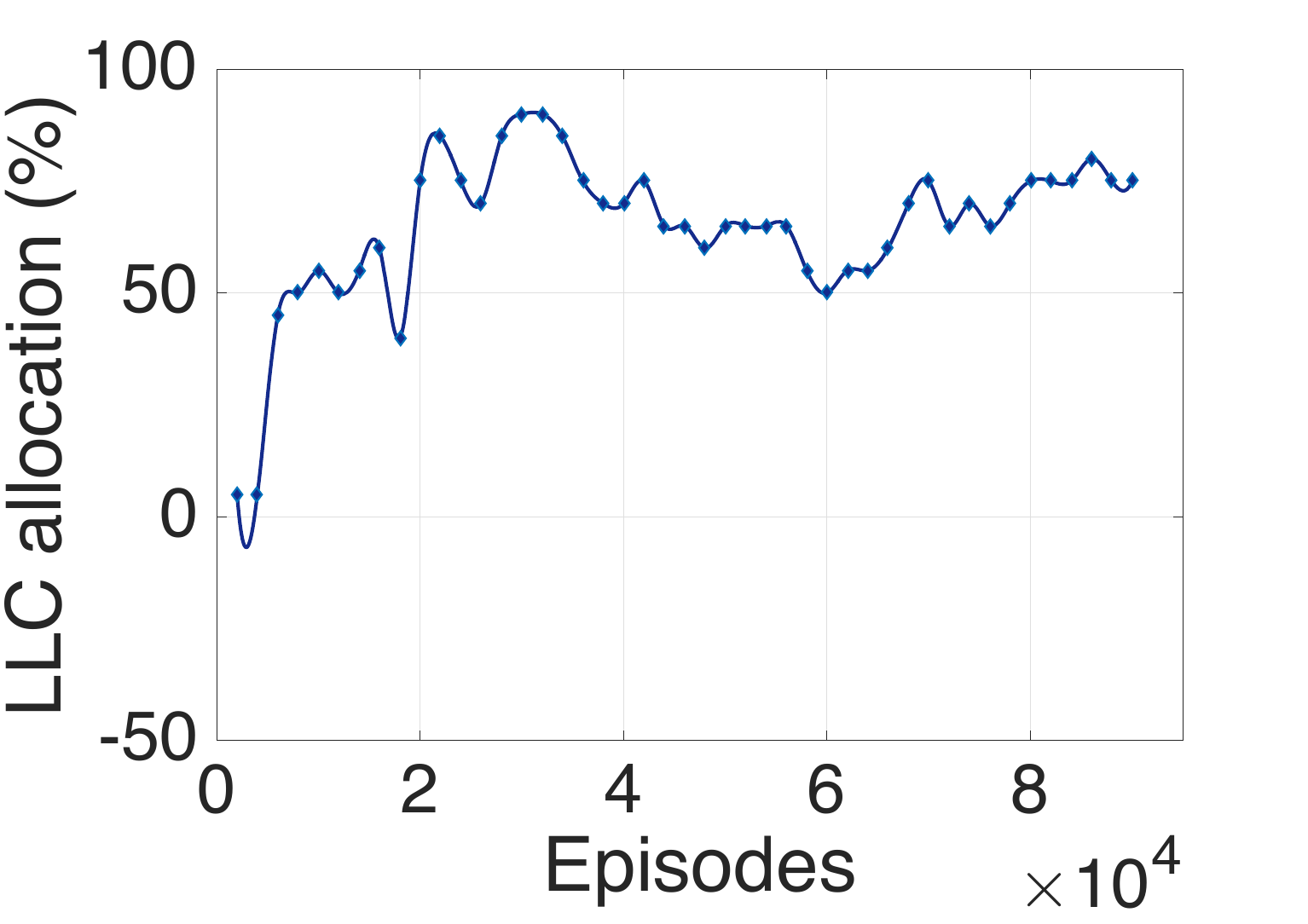}
        \subcaption{LLC allocation}
\end{subfigure}
\begin{subfigure}[t]{0.24\textwidth}
        \includegraphics[keepaspectratio=true,width=40mm]{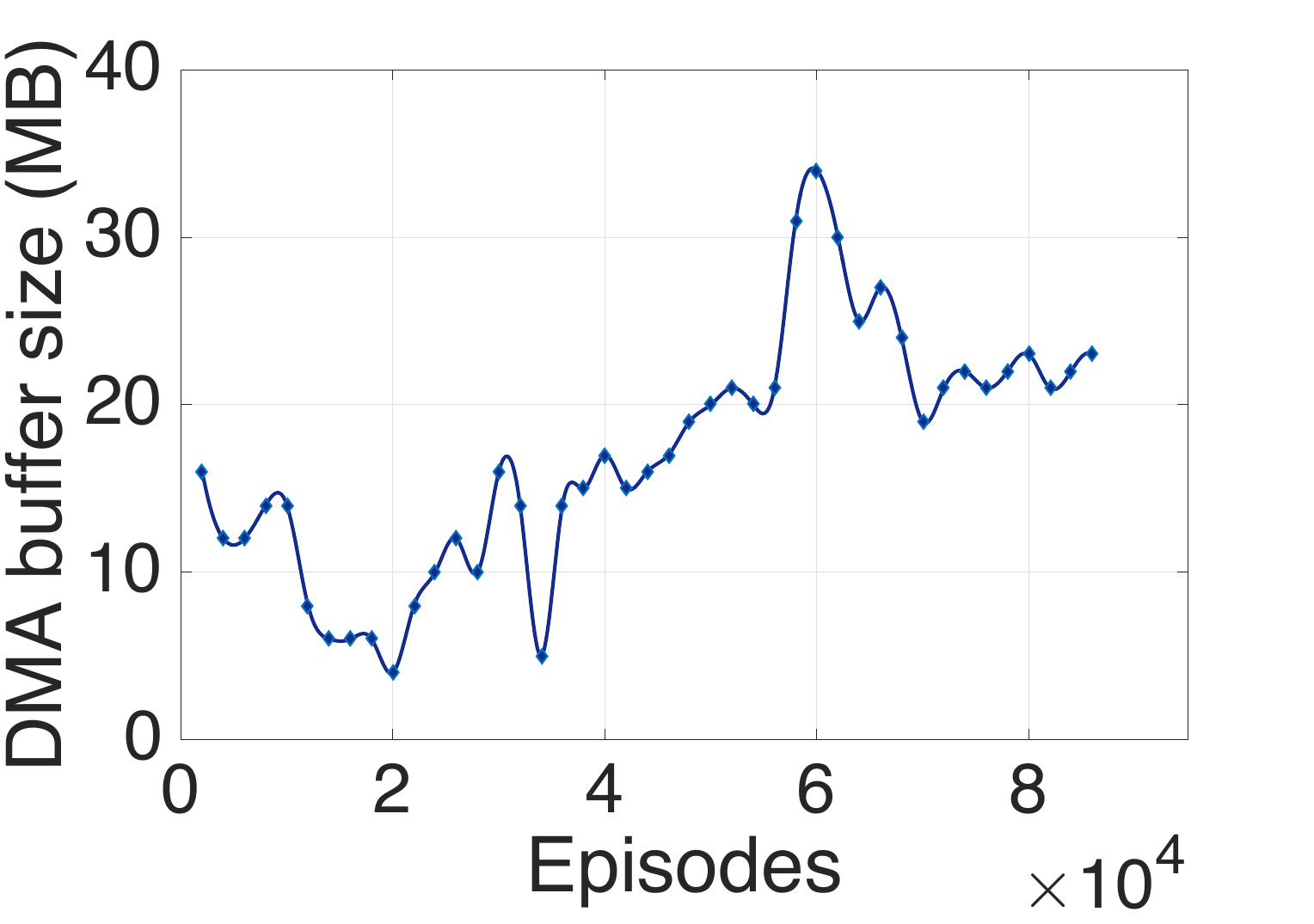}
        \caption{DMA buffer size}
\end{subfigure}
\begin{subfigure}[t]{0.24\textwidth}
        \includegraphics[keepaspectratio=true,width=40mm]{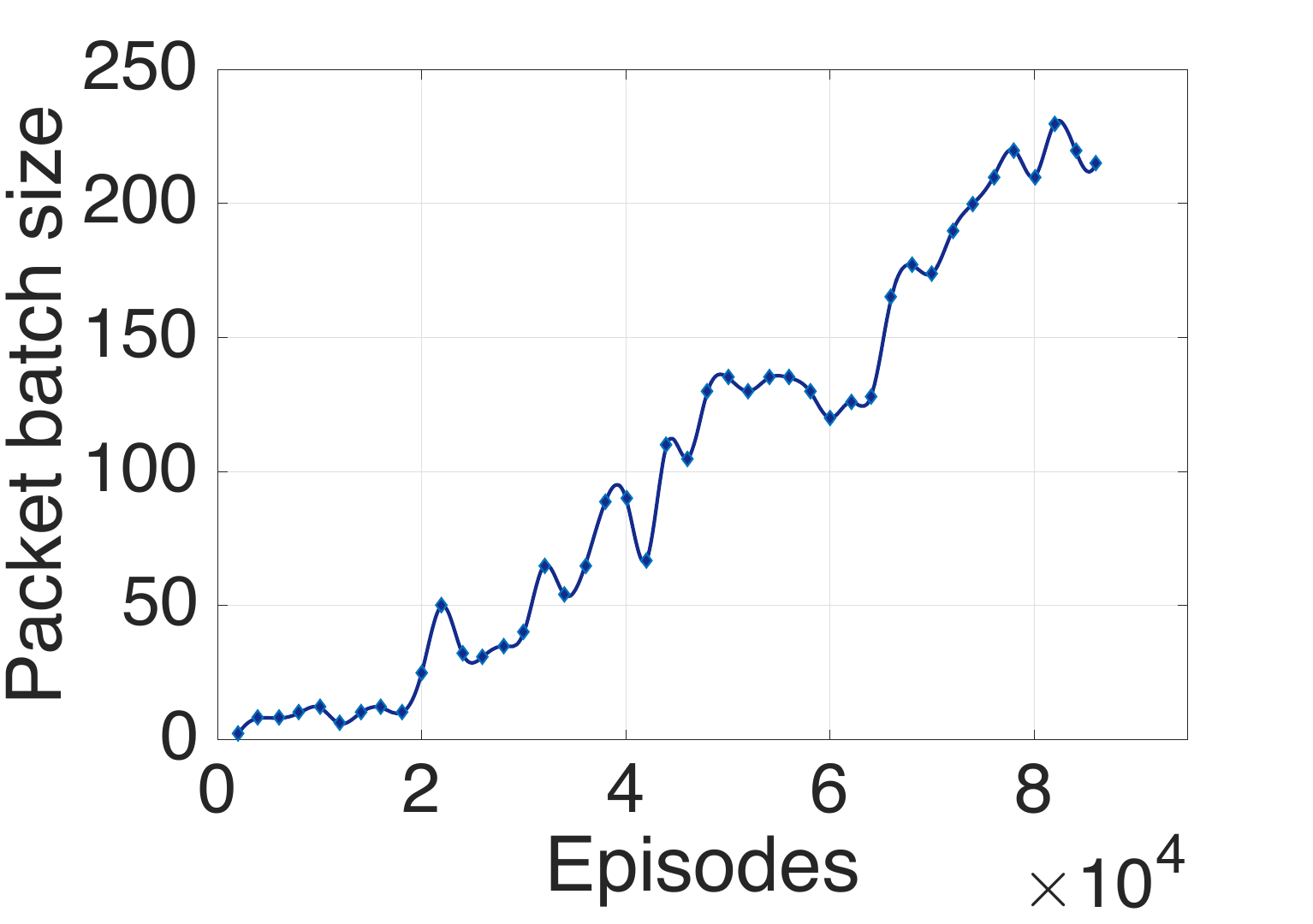}
        \caption{Packet batch size}
\end{subfigure}
     \caption{Training progress of the proposed reinforcement learning algorithm during the testing of the Energy-Efficiency SLA.}
     \label{fig:SLA_EE_results}
     \end{centering}
 \end{figure*}

 The CPU utilization and core frequency level have a significant impact on the achieved throughput and total energy consumption. For this reason, these two knobs are used to balance throughput and energy consumption. In the experiments, the core frequencies are initially set to a minimum level (1.3 GHz) to ensure the energy constraint is preserved. Later, the model tries to increase the frequency to higher values to increase the performance. As soon as it hits the constraint, it steps back (Figure~\ref{fig:SLA_maxth_results}(d)). At the same time, the model balances the CPU utilization percentage by decreasing it to 60\% to keep the power consumption under control. Throughout the training, the model tries to learn the optimal frequency setting and the correct amount of CPU allocation for increased performance and controlled energy consumption (Figure~\ref{fig:SLA_maxth_results}(a-b)).  CPU allocation and frequency settings are tuned down to decrease the consumed energy, the other three knobs (packet batch size, LLC allocation, and DMA buffer size) need to be tuned up to increase the performance. 

\begin{figure*}[t]
\begin{centering}
\hspace{1mm}
\begin{subfigure}[t]{0.45\textwidth}
    \centering
        \includegraphics[keepaspectratio=true,width=55mm]{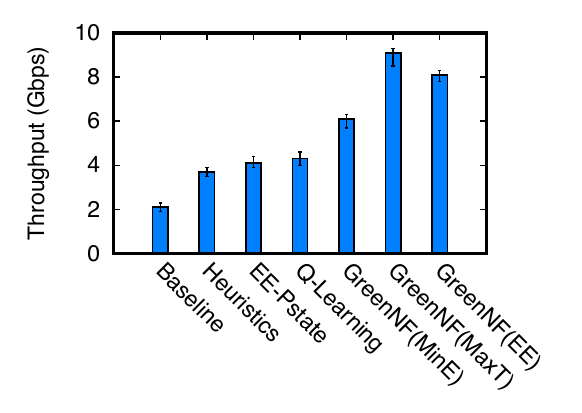}
        \caption{Achieved Throughput}
\end{subfigure}
\begin{subfigure}[t]{0.45\textwidth}
        \hspace{-5mm}
        \includegraphics[keepaspectratio=true,width=55mm]{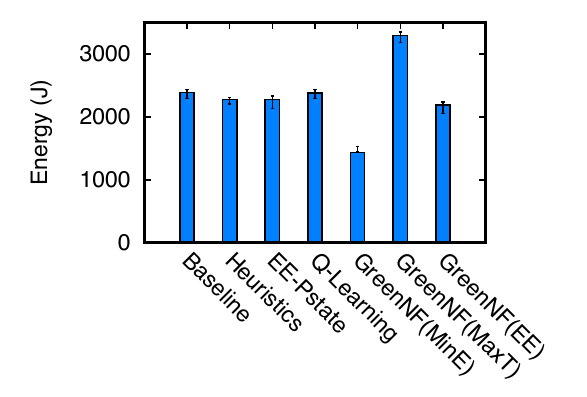}
        \caption{Energy consumption}
\end{subfigure}
 \caption{Performance comparison of different models based on throughput and energy consumption}
 \label{fig:model_comparison}
 \end{centering}
 \end{figure*}


\begin{figure*}[t]
    \begin{centering}
\begin{subfigure}[t]{0.42\textwidth}
    \centering
        \includegraphics[keepaspectratio=true,width=65mm]{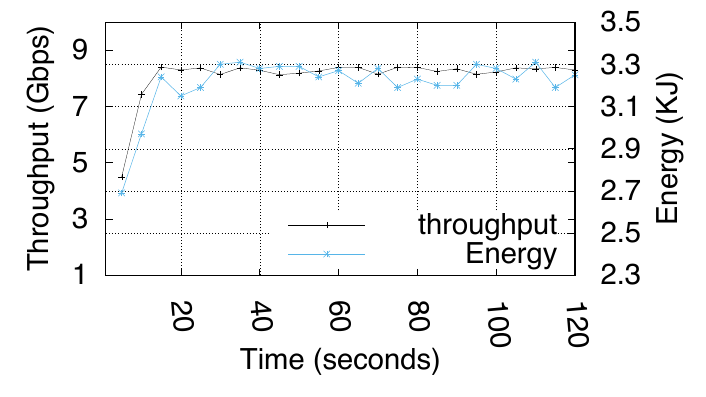}
        \caption{Achieved throughput on MaxTh with energy constraint 3.3KJ}
\end{subfigure}
\begin{subfigure}[t]{.42\textwidth}
        \includegraphics[keepaspectratio=true,width=65mm]{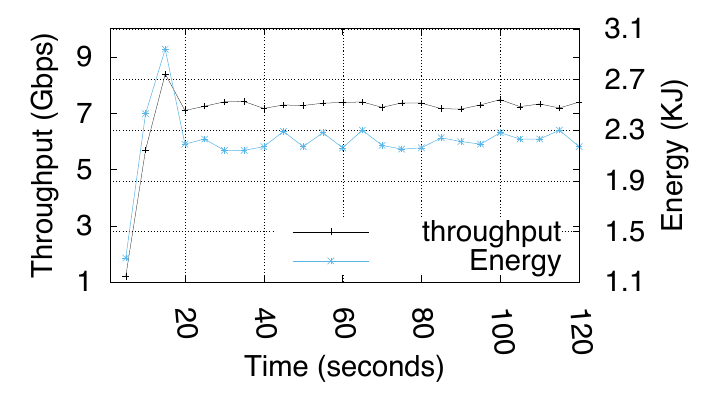}
        \caption{Energy consumption on MinE with throughput constraint of 7.5 Gbps}
\end{subfigure}
     \caption{Performance (in terms of throughput and energy consumption) of the model with different SLA's over time.}
     \label{fig:SLA_results}
     \end{centering}
 \end{figure*}
 
 We compared the result of the Maximum Throughput SLA with other existing models - baseline, Heuristics, EE-Pstate~\cite{faisal12}, and Q-learning model in Figure~\ref{fig:model_comparison}. This SLA is set to maximize the throughput.
 The baseline does not tune any system-level parameters. Therefore, the achieved throughput is the lowest compared to other models used in the experiment. The baseline heuristic algorithm tries to adjust the parameters in real time. However, it does not use any prior knowledge about the system. It makes decisions based on purely real-time feedback from the network using predefined static rules. Such decision-making is slow and takes a long time to converge. Still, the heuristic-based approach can achieve $2\times$ performance improvement over baseline. On the other hand, EE-Pstate uses thresholding on the p-state level of the processor cores and leaves other control knobs without optimization. The Q-learning-based model also has difficulty increasing the throughput. It works with predefined discrete levels of parameters. Therefore, fine-tuning the parameters is difficult in real-time. 
 We can see a $4.4\times$ throughput improvement on MaxTh SLA over the baseline model and almost $2\times$ improvement over the Heuristics, EE-Pstate, and Q-learning model. Moreover, the energy efficiency (throughput/energy) is improved by $1.5\times$ compared to the Heuristics, EE-Pstate, and Q-learning model.
 
\subsection{Minimum Energy SLA}
\label{subsec:energy}

The Minimum Energy SLA aims to minimize the energy consumption of the NF processing while maintaining a specific minimum throughput defined in the SLA. In the experiment, we set the minimum throughput constraint to $7.5$ Gbps, and if the model violates that constraint, it gets no rewards.
The model only receives rewards when it can maintain the throughput constraint, and the reward gets better when it reduces energy consumption. 

Initially, the model starts to find the control settings that can yield high throughput even with high energy consumption. This setting is better than any setting that fails to maintain the throughput guarantee. As the training progresses, the model tries to find a solution that reduces energy consumption while maintaining the throughput constraint. 
The model effectively learns to maintain the throughput constraint. Figures~\ref{fig:SLA_minE_results}(c-g) shows how the model tunes the control knobs to achieve this during training.
The model starts with a high CPU utilization and high core frequency value (Figures ~\ref{fig:SLA_minE_results}(c-d))). The packet batch size is also large to facilitate high throughput. 
The model tries to find settings that can reduce energy consumption to receive more rewards. It tries to keep the LLC allocation stable and increases both batch size and buffer size (Figures ~\ref{fig:SLA_minE_results}(e-g)). It reduces the per-packet energy cost; however, the reduction is not very significant. At the end of the training period, it tries to decrease CPU utilization and keeps the core frequency high. Simultaneously, the model increases the LLC and buffer size to compensate for the decreased CPU. This strategy reduces energy consumption and keeps the throughput above the set constraint by the SLA.

We compare the performance of Minimum Energy (MinE) SLA with the baseline algorithm, Heuristics-based approach, EE-Pstate.
Minimum Energy SLA can achieve $3\times$ higher performance than baseline, also more than 30\% improvement over Heuristics, EE-Pstate, and Q-learning while reducing the energy consumption by almost 60\% compared to other models.

\subsection{Energy-efficiency SLA}
\label{subsec:energy-efficiency}
In this experiment, we test the Energy Efficiency SLA and its progress over the training period. This is an unconstrained optimization of SLA where the model can receive more rewards when achieving high energy efficiency. The model starts with a controlled setting of low energy efficiency and then tries to increase the reward by searching for more energy-efficient settings.

Figure~\ref{fig:SLA_EE_results} records the throughput, energy consumption, efficiency, and status of the control knobs during the training process. During the training process, the model selects a setting with low energy efficiency (Figure~\ref{fig:SLA_EE_results}(c)). The initial throughput (Figure~\ref{fig:SLA_EE_results}(a)) is around 4 Gbps with an energy consumption level of around 4200 Joules (Figure~\ref{fig:SLA_EE_results}(b)). Then the model tries to find the settings that consume less energy with similar performance. This policy increases energy efficiency. Then around the 12000th episode, a significant drop in the throughput and energy consumption is observed; however, the efficiency remains the same. The model can also achieve high efficiency when it processes fewer packets using less energy. The model explores that region and then increases its throughput again. Then we can see the model tries to keep the energy consumption level stable and increases the throughput, which also results in high energy efficiency. Then, around the 16000th episode, the model attempts to stabilize the energy and increases the throughput with some oscillations. To perform that, the model increases the CPU allocation, CPU frequency, and the LLC allocation (Figures~\ref{fig:SLA_EE_results}(d,e-f)), but decreases the DMA size. As a result, a spike in energy efficiency is observed until the 40000th episode. Afterward, the model changes the policy to reduce CPU allocation sharply. In the meantime, the model steadily increases the batch size and the DMA buffer size. This keeps the throughput steady at around 6 Gbps; however, it decreases energy consumption. As a result, we observe a steady increase in energy efficiency. After the 60000th episode, we observe a steady allocation of CPU around 200\%, and the model tries to increase the CPU frequency to the maximum level.
Meanwhile, an increased batch size helps to achieve high throughput. The model also stabilizes the DMA buffer size. Initially, the model tries to increase throughput by increasing the CPU allocation—however, the increased energy consumption is a barrier to achieving higher efficiency. Then, the model decreases the CPU allocation and tries to find optimal settings to increase energy efficiency. Figure~\ref{fig:model_comparison} shows that the Energy Efficient SLA - GreenNFV(EE), can achieve almost $4\times$ performance improvement over the baseline and $2\times$ improvement over the Heuristics, EE-Pstate, and Q-learning models while consuming energy similar to these models.

We also tested our SLA-based models with fixed SLA constraints. Maximum throughput SLA is fixed with energy constraint 3.3KJ. Initially, the model tries to achieve the maximum throughput that the energy constraint allows. The model tries to allocate resources so that the energy constraint violation does not occur. Initially, we observe SLA violations and oscillations in performance and energy consumption. Afterward, the model settles down at around 8 Gbps throughput without violating the energy constraint. 
Minimum Energy SLA is fixed with a throughput constraint of 7 Gbps. That means the Minimum Energy SLA should provide at least 7 Gbps performance while minimizing the energy. Figure~\ref{fig:SLA_results} shows that initially, MinE SLA takes some time to reach the SLA goal. We can see an overshoot in performance with high energy consumption. Finally, the model reduces the throughput to 7 Gbps and also reduces energy consumption.  

\begin{figure}[t]
\centering
\includegraphics[keepaspectratio=true,angle=0,width=50mm]{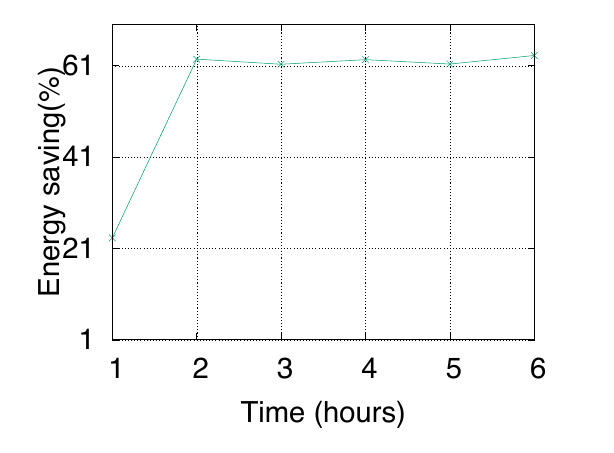}
\caption{Total energy consumption (including the energy cost of training RL algorithm) improvement compared to other models} 
\label{fig:energy_improvement}
\end{figure}

The RL model itself consumes energy during the training process. However, the GreenNFV model needs to be trained only once before deployment and is run many times during the decision-making process. Once the model is trained, the decision is obtained in constant time. The initial training cost is amortized over many subsequent future decision-making runs. To see the impact we calculated the Energy saving, $E_s$, as follows:
\begin{equation}
    E_s = (E_{nf} + E_t - E_b) / (E_{nf} + E_t)
\label{eq:energy_improvement}
\end{equation}

where $E_{nf}$ is the energy consumption of the Nfs, $E_t$ is the energy consumption during the training, and $E_b$ is the energy consumption of the baseline model. 
Figure~\ref{fig:energy_improvement}(c) shows that the MinE algorithm can consume 23\% less energy even when the energy cost of the model training process is included and over time it reaches 62\%.

%% file: section-conclusion.tex
\section{Conclusion}
\label{sec:conclusion}

This study explores a novel energy optimization framework for the Network Function Virtualization (NFV) platform with various user-defined service level agreements (SLAs). Existing literature contains different system-level optimization techniques to increase the processing speed of the Network Function (NF) chains. In our study, we extensively explored the impact of energy consumption without impacting performance. Our novel solution, called \name, employs a Deep Reinforcement Learning (DRL) based approach by translating the resource scheduling problem into a deep deterministic policy gradient (DDPG) algorithm, a value-based actor-critic reinforcement learning algorithm, which is very effective for continuous (real-valued) and high-dimensional action space. 
We present three novel resource optimization models based on different energy-aware service level agreements
(SLAs), which enable the telecommunication service providers (TSPs) to minimize energy consumption without compromising the performance guarantees given to the customers. \name shows $4.4\times$ performance improvement over the baseline settings while consuming 33\% less energy with the {\em Throughput SLA}. When used with the {\em Energy SLA}, it can achieve $3\times$ throughput improvement while reducing energy consumption by 50\%.  

In future work, we plan to incorporate software-defined networking (SDN) and NF controllers to provide higher flexibility. We envision a model where both the SDN controller and NF controller can update each other to perform more effective flow scheduling.